# Dispersion models describing coupled systems in doped crystals. II. Infrared phonon excitations in GaAs single crystals and phonon–plasmon coupling in Zn-doped GaAs

Daniel Franta,[1,2,a)] 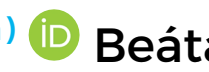 Beáta Hroncová,[1,3] 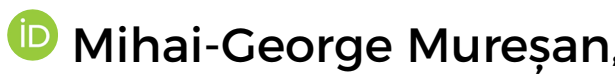 Mihai-George Mureșan,[4] 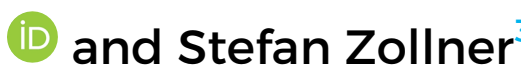 and Stefan Zollner[3]

## AFFILIATIONS

[1] Department of Plasma Physics and Technology, Faculty of Science, Masaryk University, Kotlářská 2, 611 37 Brno, Czechia
[2] Central European Institute of Technology CEITEC, Brno University of Technology, Purkyňova 123, 612 00 Brno, Czechia
[3] Department of Physics, New Mexico State University, MSC 3D, P.O. Box 30001, Las Cruces, New Mexico 88003-8001, USA
[4] HiLASE Centre, Institute of Physics, Academy of Sciences of the Czech Republic, Za Radnicí 828, 252 41 Dolní Břežany, Czechia

**Note:** This paper is part of the Special Topic on Advances in Spectroscopic Ellipsometry Methods and Materials Characterization.

[a)] **Author to whom correspondence should be addressed:** franta@physics.muni.cz

## ABSTRACT

We present a temperature-dependent infrared study of intrinsic and Zn-doped GaAs single crystals analyzed within a recently developed dispersion framework for coupled phonon–carrier systems. Reflectance, transmittance, and ellipsometry data (300–440 K) were fitted simultaneously using response functions that incorporate Gaussian-broadened dipole distributions for single-phonon excitations and triangle–delta–triangle representations for multi-phonon absorption. The approach enables a clear separation of lattice and free-carrier contributions and reveals a pronounced Fano-type asymmetry of the TO-phonon resonance in heavily doped samples, originating from the interference between lattice vibrations and the Drude background of light and heavy holes. The model provides physically consistent optical constants across the far- and mid-infrared, showing improved internal agreement compared with existing datasets. These results demonstrate that the extended dispersion formalism yields an accurate description of single-phonon, multi-phonon, and free-carrier processes in GaAs and establishes a validated basis for quantitative modeling of GaAs-based photonic and optoelectronic structures.

## I. INTRODUCTION

Gallium arsenide (GaAs) single crystals have played a key role in photonic and optoelectronic technologies for several decades. Owing to their direct bandgap, high carrier mobility, and excellent infrared transparency, GaAs-based structures are widely used in light-emitting diodes, high-speed electronics, and especially in semiconductor lasers.[1] In high-power and diode-pumped solid-state laser (DPSSL) systems, GaAs serves both as the active medium in pump diodes and as a constituent of advanced optical components. Accurate knowledge of its temperature-dependent optical properties is, therefore, essential for predicting device behavior under thermal load.

GaAs laser diodes exhibit a pronounced sensitivity to temperature: their bandgap energy narrows with increasing temperature, causing a red shift of the emission wavelength, while the refractive index increases by approximately $2.7 \times 10^{-4}\,\mathrm{K}^{-1}$ near $1\,\mu\mathrm{m}$.[2] These effects can detune pump diodes from the absorption band of the gain medium and induce thermal lensing in downstream optics. In vacuum or space environments, where convective cooling is absent, local overheating may become severe and accelerate defect formation or facet degradation.[3] Consequently, thermal modeling of GaAs components is critical for ensuring long-term stability and performance of DPSSL systems.

Beyond semiconductor pump sources, GaAs and related GaAs/AlGaAs heterostructures are increasingly used in high-precision optical elements, including high-reflectivity crystalline "supermirror" coatings[4] for laser cavities. These epitaxial multilayers exhibit extremely low optical absorption and thermal noise, and their thermal conductivity significantly exceeding that of conventional dielectric coatings. Such properties make crystalline GaAs/AlGaAs mirrors attractive for high-power and metrology-grade laser systems, where accurate characterization of heat deposition, refractive-index changes, and thermally induced strain is indispensable. A reliable set of temperature-dependent optical constants is, thus, a prerequisite for predictive thermo-optical simulations of these advanced optical elements.

GaAs space solar cells offer high efficiency and exceptional radiation tolerance, making them the preferred photovoltaic technology for satellites and deep-space missions.[5] Their performance, however, is tightly linked to the temperature dependence of GaAs optical properties, since variations in bandgap energy and refractive index directly affect absorption and carrier generation. As temperature increases, the bandgap narrows and the absorption edge shifts to longer wavelengths, modifying the spectral response and overall power conversion efficiency.

The aim of this work is to develop an advanced temperature-dependent dispersion model for intrinsic and doped GaAs crystals in the far-infrared spectral range. This study serves as a pilot validation of newly proposed dispersion models that describe coupled bound and free charge carriers and enable reconstruction of optical constants over the widest feasible spectral and temperature intervals. Our objective is not to provide a fundamental reinterpretation of the optical properties of GaAs—these are well established and comprehensively summarized, for instance, in the review by Blakemore.[6] Instead, the focus is on constructing and verifying a physically consistent framework capable of modeling the individual absorption processes in detail.

This work builds on our previous development of an advanced temperature-dependent dispersion model for crystalline silicon,[7,8] which has matured into a robust generator of optical constants for silicon wafers routinely used as reference substrates in thin-film optics. Many functional materials and device structures, however, cannot be grown on silicon and require other substrates such as GaAs. The long-term motivation of this research is, therefore, to extend the same level of accuracy and reliability achieved for silicon to GaAs wafers, enabling their use as well-characterized reference substrates in optical metrology and facilitating precise modeling of GaAs-based photonic components across a broad range of applications.

Although reliable optical constants of undoped GaAs are available in the visible and ultraviolet spectral regions,[9–12] the situation in the infrared is markedly different. Existing datasets in the far- and mid-infrared often lack the consistency, temperature coverage, or physical completeness needed for predictive modeling. As a by-product of this work, we provide a set of temperature-dependent infrared optical constants of GaAs in the range 300–440 K that meets these requirements and enables quantitatively reliable thermo-optical simulations.

## II. SAMPLES

For optical characterization, three GaAs samples (manufacturer CMK[13]) grown by the vertical gradient freezing (VGF) method with (100) surfaces were used. Sample #1 is undoped GaAs in the form of a double-sided polished wafer with a nominal thickness of $625 \pm 25\,\mu$m. The reported resistivity of this sample falls within the range $(1.61 - 2.23) \times 10^8\,\Omega\,$cm.

Samples #2 and #3 are Zn-doped GaAs wafers, with basic parameters listed in Table I. Sample #2 is a double-sided polished wafer with a nominal thickness of $350 \pm 25\,\mu$m, whereas sample #3 is a single-sided polished wafer with a nominal thickness of $450 \pm 25\,\mu$m.

## III. EXPERIMENTAL DATA AND DATA PROCESSING

Experimental data were acquired using a J. A. Woollam IR-VASE ellipsometer and two Bruker IR spectrophotometers. The Vertex 70v spectrophotometer, equipped with a mercury lamp, extended the spectral range into the far-IR region but did not allow for stabilized sample temperature control. The Vertex 80v spectrophotometer was equipped with temperature-controlled sample holders, with the temperature maintained using an Instec mK1000 precision controller. An overview of all measurements is provided in Table II. The nominal spectral resolution for all measurements was $1\,\text{cm}^{-1}$.

Our experimental data were complemented by tabulated values of the thermal expansion coefficient $\alpha$ and the refractive index $n$ in the near- and mid-IR regions obtained from the literature (see Table II).

The complete set of all experimental and tabulated data listed in Table II was fitted using the newAD2 software package,[19] which implements the balanced Levenberg–Marquardt algorithm and enables the processing of heterogeneous data sets.[20] This procedure yielded the temperature-dependent optical constants and the parameters of the dispersion model presented in Secs. IV A–IV E. The quoted parameter uncertainties (given in parentheses after the corresponding parameter values) represent the statistical errors obtained from the regression procedure, reflecting the local curvature of the merit function at the optimum. They do not constitute a full metrological uncertainty analysis and should, therefore, not be interpreted as absolute physical accuracies.

**TABLE I.** List of Zn-doped GaAs samples used for optical characterization. The first two rows correspond to the nominal values provided by the manufacturer. Static conductivity $\sigma_0$ and resistivity $\rho_0 = 1/\sigma_0$ correspond to the optical conductivity $\hat{\sigma}(\omega)$ determined in this work at low frequencies.

| | #2 | #3 |
|---|---|---|
| Concentration ($\text{cm}^{-3}$) | (2.3–8.03)$\times10^{17}$ | (0.7–2)$\times10^{19}$ |
| Resistivity ($\Omega$ cm) | 0.116–0.0401 | >0.001 |
| St. cond. $\sigma_0$ (kS/m) | 1.464 | 15.71 |
| St. resist. $\rho_0$ ($\Omega$ cm) | 0.0683 | 0.006 37 |

**TABLE II.** List of experimental and tabulated data fitted simultaneously using the dispersion model (1) in our optical characterization. The nominal spectral resolution for all measurements was 1 cm$^{-1}$.

| Instrument | Quantity | Sample # | Range (cm$^{-1}$) | Angle of inc. (deg) | Temperature (K) | Figures |
|---|---|---|---|---|---|---|
| Woollam IR-VASE | $I_s$, $I_c$, $I_n$ | 1, 2, 3 | 250–6000 | 55, 65, 75 | Laboratory | 7 |
| Bruker Vertex80v | $R$ | 1, 2, 3 | 70–680 | 10 | 305, 325, 350, 380, 410, 440 | 5, 7 |
| Bruker Vertex80v | $T$ | 1 | 70–680 | 0 | 310, 325, 350, 375, 400, 430 | 14 |
| Bruker Vertex70v | $R$ | 1, 2, 3 | 25–680 | 10 | Laboratory | 6, 7, 14 |
| Bruker Vertex70v | $T$ | 1 | 25–680 | 0 | Laboratory | 14 |
| Reference | Quantity | | Range (eV) | | Temperature (K) | Figure |
| Novikova[14] | $\alpha$ | | | | 28–348 | 1 |
| Glazov–Pashinkin[15] | $\alpha$ | | | | 392–1185 | 1 |
| Marple[16] | $n$ | | 0.73–1.47 | | 103, 187, 300 | 2 |
| Hambleton *et al.*[17] | $n$ | | 0.25–0.65 | | Laboratory | 2 |
| Skauli *et al.*[18] | $n$ | | 0.6–1.28 | | 294.15 | 2 |

## IV. DISPERSION MODEL AND RESULTS

The spectral dependence of the response function of a GaAs single crystal in the IR region exhibits single dielectrically active TO-phonon absorption. Accordingly, the single-phonon response can be approximated in a first approach by the Lorentz model of a damped harmonic oscillator. For free carriers in doped semiconductors, the standard description is provided by the classical Drude model. Thus, the Drude–Lorentz model represents the classical model of doped semiconductors when coupling is neglected.

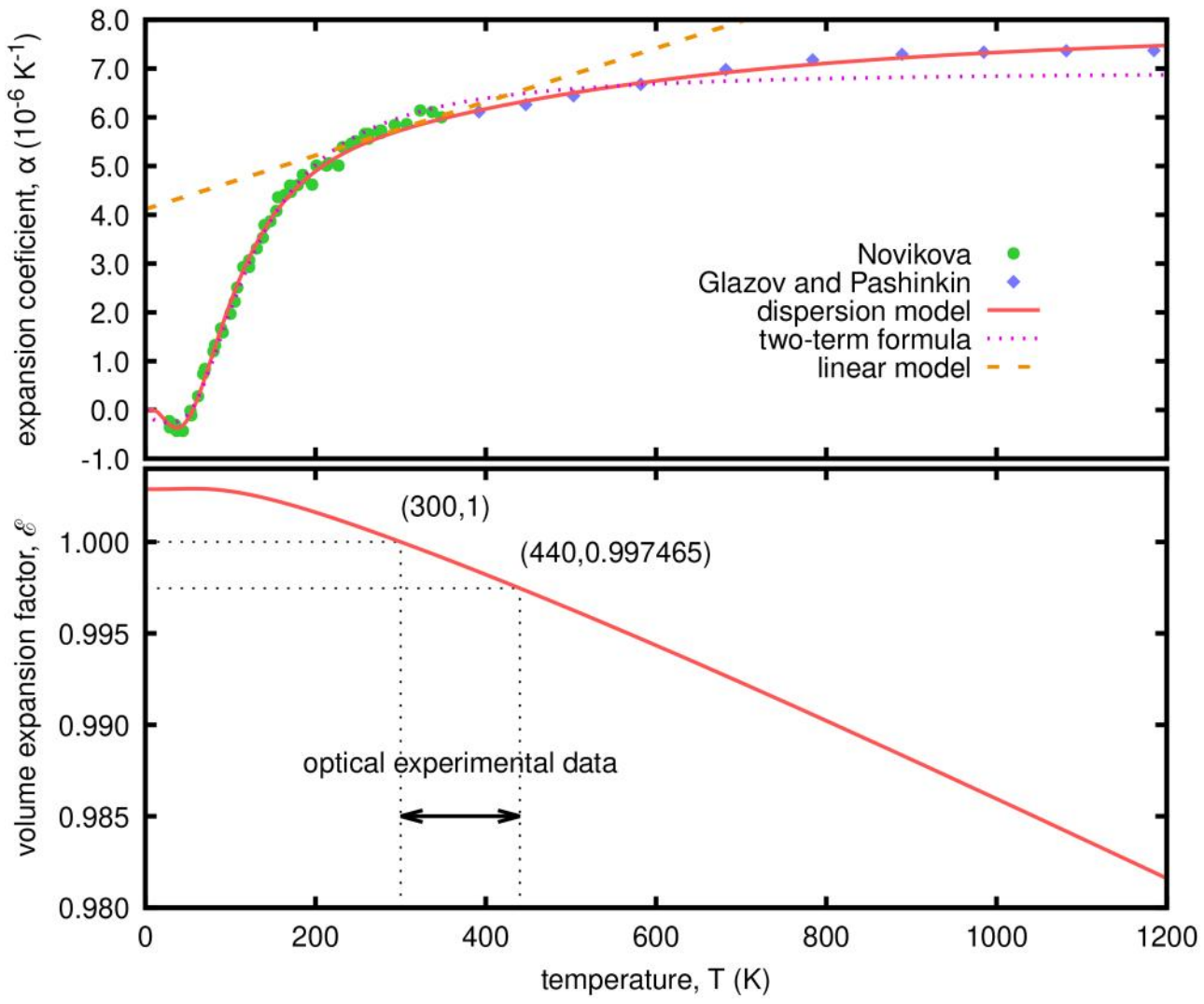

**FIG. 1.** (a) Temperature dependence of the thermal expansion coefficient $\alpha(T) = \partial e(T)/\partial T$. Experimental data are taken from the literature.[14,15] The theoretical curve used in the dispersion model is described by the three-term formula (5) with parameters listed in Table III. For comparison, the fit of $\alpha(T)$ obtained using the two-term formula is also shown. The linear approximation is based on data from Reeber and Wang.[31] (b) Volume expansion factor $\mathcal{E}(T)$ of GaAs calculated according to (3).

In coupled systems, the coupling manifests as an asymmetric absorption peak. Therefore, instead of the Drude–Lorentz model, one can employ the factorized oscillator model for conductive materials, known as the Kukharskii model.[21,22]

As shown previously, this model can produce an unphysical response function due to incorrect asymptotic behavior.[23,24] To address this deficiency, constraints on the transverse and longitudinal broadening parameters are necessary. These constraints are analogous to those used in the factorized oscillator model for non-conducting materials.[25] With these constraints, the Kukharskii model corresponds to the response of classical coupled particles.[24] Without applying the constraints, the model fails at higher energies and cannot relate the response function to the finite density of charged particles in the system, as the sum rule integral diverges.

For the interpretation of experimental spectrophotometric and ellipsometric data, it is essential to move beyond classical models using Lorentzian broadening because these models are insufficient for accurately describing the optical constants over a wide spectral range. Additionally, to account for temperature dependence, dispersion models must be linked to sum rules that relate the density of matter to excitation strengths.[26–28]

In the present dispersion model, the spectral- and temperature-dependent dielectric function of GaAs is decomposed into four contributions describing the susceptibilities of electron valence-to-conduction band transitions (el), single-phonon excitations (ph), multi-phonon excitations (mph), and free carriers (fc), respectively,

$$\hat{\varepsilon}(E, T) = 1 + [\hat{\chi}_{\mathrm{el}}(E, T) + \hat{\chi}_{\mathrm{ph}}(E, T) + \hat{\chi}_{\mathrm{mph}}(E, T) + \hat{\chi}_{\mathrm{fc}}(E, T)]\,\mathcal{E}(T). \quad (1)$$

Here, $\mathcal{E}(T)$ is the volume expansion factor, which is independent of the photon energy $E$.

In our nomenclature, a hat above each symbol indicates a complex quantity. Furthermore, the real and imaginary parts of a complex quantity are denoted by the first subscript of the corresponding symbol, e.g., $\hat{\varepsilon}(E, T) = \varepsilon_r(E, T) + i\varepsilon_i(E, T)$. An exception is the complex refractive index, which is written in the conventional form $\hat{n} = n + \mathrm{i}k$.

In the following, the complex functions $\hat{\chi}_t(E, T)$ are referred to as the susceptibilities of individual contributions, although in reality, they should also include the volume expansion factor $\mathcal{E}(T)$.

Then, for the total f-sum rule and the partial f-sum rule, it is necessary to write[27]

$$\begin{aligned}\int_0^\infty E\,\varepsilon_\mathrm{i}(E, T)\,\mathrm{d}E &= \mathcal{E}(T)N,\\ \int_0^\infty \mathcal{E}(T)\,E\,\chi_{\mathrm{i},t}(E, T)\,\mathrm{d}E &= \mathcal{E}(T)N_t(T),\\ N &= \sum_t N_t(T),\end{aligned} \tag{2}$$

where $t = $ el, ph, mph, fc and $N$ is the total transition strength of all charged particles in the studied system at the reference temperature. The partial transition strengths $N_t(T)$ may, in general, depend on temperature. For example, in the case of multi-phonon excitations, the strengths depend on Bose–Einstein statistical factors,[29,30] while the strengths of single-phonon excitations do not depend on temperature because the increase in probabilities of absorption and stimulated emission with temperature exactly cancel in their net contribution.[29,30]

## A. Thermal expansion

Part of the dispersion formula is the linear thermal expansion function $e(T)$. The volume expansion factor is calculated from the linear thermal expansion as

$$\mathcal{E}(T) = [1 + e(T)]^{-3}. \tag{3}$$

These functions describe how the density of charged particles in the studied system and its linear size, e.g., the substrate thickness $d_\mathrm{s}(T)$, change with temperature $T$,

$$d_\mathrm{s}(T) = d_\mathrm{s}(T_\mathrm{R})[1 + e(T)], \tag{4}$$

where $T_\mathrm{R}$ is the reference temperature, taken here as $T_\mathrm{R} = 300\,\mathrm{K}$. Experimentally, it is easier to obtain the coefficient of thermal expansion $\alpha(T)$, which is the temperature derivative of the linear expansion.

In the present dispersion model, linear thermal expansion is described using three average Bose–Einstein statistical factors,[8,27,31–33]

$$e(T) = \sum_{j=1}^{3}\left[\frac{e_j}{\exp(\Theta_j/T) - 1} - \frac{e_j}{\exp(\Theta_j/T_\mathrm{R}) - 1}\right], \tag{5}$$

where $e_j$ and $\Theta_j$ are dispersion parameters. Parameter $\Theta_j$ represents the effective phonon energy $E_j$ (or phonon frequency $f_j$), expressed in temperature units via the Boltzmann relation ($k_\mathrm{B}\Theta_j = E_j = hf_j$). These parameters are determined together with other dispersion parameters during the processing of the heterogeneous data set (see Table II). The fit of the thermal expansion coefficient $\alpha$ published by Novikova,[14] Glazov and Pashinkin[15] is shown in Fig. 1(a) and Table III, while Fig. 1(b) shows the corresponding volume expansion factor.

**TABLE III.** Dispersion parameters describing thermal expansion.

| $j$ | $e_j$ | $\Theta_j$ (K) | $f_j$ (THz) |
|---|---|---|---|
| 1 | −0.000 064(3) | 92(2) | 1.924 |
| 2 | 0.002 460(3) | 346.3(7) | 7.216 |
| 3 | 0.003 199(11) | 2203(8) | 45.89 |

By comparing the $\Theta_j$ parameters with the phonon frequencies of GaAs[6] in the range 0–8.55 THz, it is evident that $\Theta_1$ corresponds to effective acoustic phonon energies and $\Theta_2$ to the effective optical phonon energies. The parameter $\Theta_3$ represents a phenomenological correction term. Although it does not have a clear microscopic physical interpretation, the third term is essential for an accurate description of the thermal expansion coefficient over a wide temperature range, as demonstrated in Fig. 1(a).

## B. Electron valence-to-conduction band transitions

The refractive index of materials and its dispersion in the transparency region are primarily determined by electronic excitations from the occupied valence band to the unoccupied conduction band. In the far-IR region, the contribution of these excitations is often approximated by a constant $\varepsilon_\infty - 1$, which is insufficiently accurate when interpreting interference patterns in double-side polished substrates. To describe the valence electron contribution to the response function more accurately, the dispersion model of harmonic oscillators (Sellmeier formula) with three terms (poles) can be employed,

$$\hat{\chi}_{\rm el}(E,T) = \frac{2}{\pi}\sum_{j=1}^{3}\frac{N_{{\rm el},j}}{E^2_{{\rm el},j}(T) - E^2} + \frac{{\rm i}N_{{\rm el},j}}{E_{{\rm el},j}(T)}\left[\delta(E - E_{{\rm el},j}(T)) - \delta(E + E_{{\rm el},j}(T))\right], \quad (6)$$

where $N_{{\rm el},j}$ is the effective transition strength of the individual pole at energy $E_{{\rm el},j}(T)$. In practice, the imaginary part is omitted in the implementation of the dispersion model.

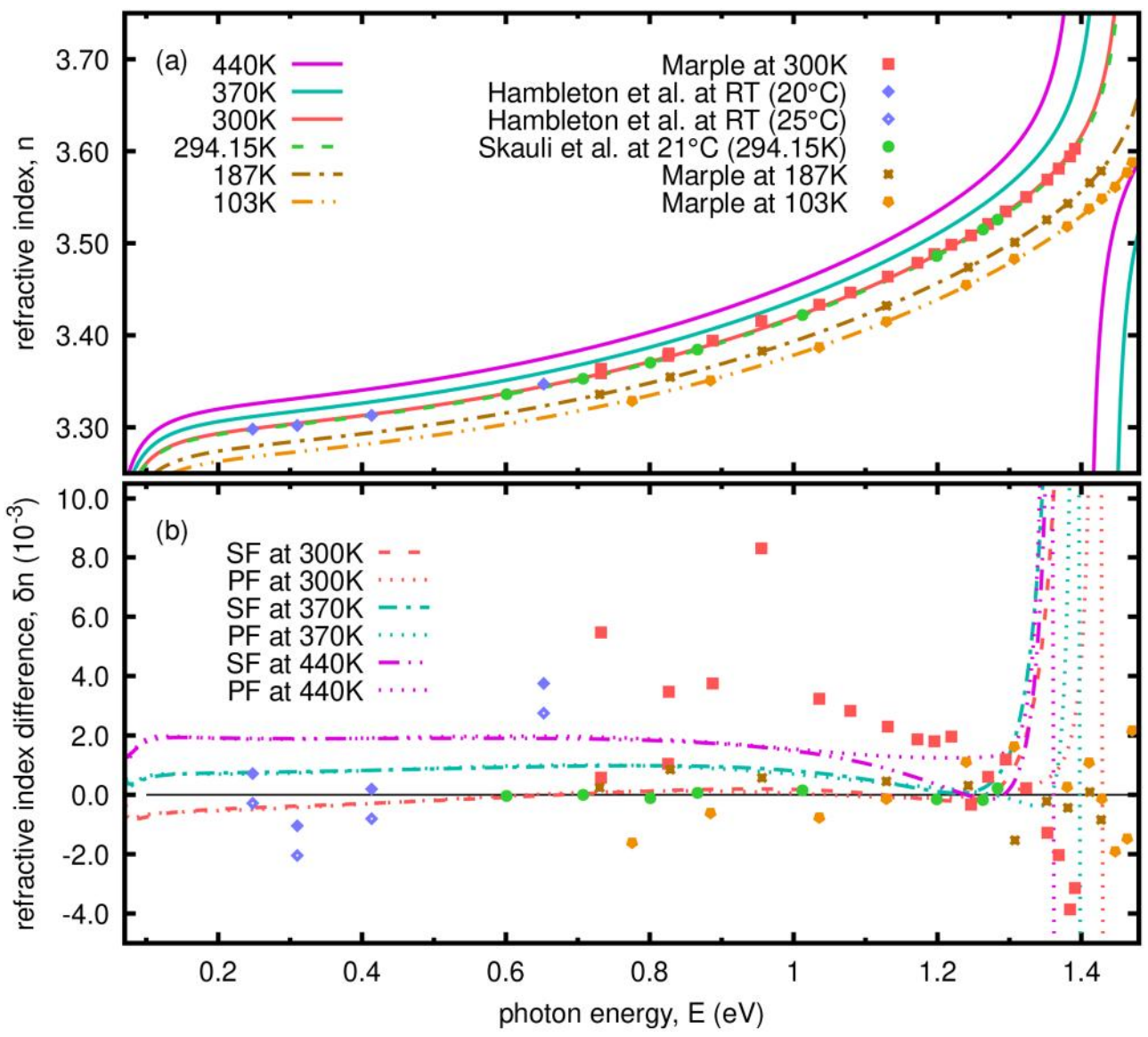

**FIG. 2.** (a) Spectral dependence of the refractive index $n$ of undoped GaAs, $n(E, T) + \mathrm{i}k(E, T) = \sqrt{\hat{\varepsilon}(E, T)}$, in the transparency region, calculated for different temperatures using the dispersion model (1). Symbols represent tabulated experimental data obtained by the minimal deviation method (MDM) by Marple[16] and Hambleton *et al.*[17] The SHG data published by Skauli *et al.*[18] are also included as symbols. (b) Differences of the refractive index with respect to the present dispersion model. The curves show the differences between refractive indices calculated using the formulas reported by Skauli *et al.*[18] (SF—Sellmeier formula; PF—Pikhtin formula) and the dispersion model (1). The symbols represent the differences between the experimental data shown in panel (a) and the corresponding values calculated using the dispersion model (1).

The positions of the poles are temperature dependent according to a Bose–Einstein statistical factor[7,8,27,34,35]

$$E_{{\rm el},j}(T) = E^{300\,{\rm K}}_{{\rm el},j}\left[\varepsilon^{0\,{\rm K}}_{{\rm el},j} + \left(1 - \varepsilon^{0\,{\rm K}}_{{\rm el},j}\right)\frac{\exp(\Theta/T_{\rm R}) - 1}{\exp(\Theta/T) - 1}\right], \quad (7)$$

where $E^{300\,{\rm K}}_{{\rm el},j}$ is the pole position at the reference temperature. The parameter $\varepsilon^{0\,{\rm K}}_{{\rm el},j}$ determines the relative position of the poles at 0 K with respect to the reference temperature

$$\varepsilon^{0\,{\rm K}}_{{\rm el},j} = \frac{E^{0\,{\rm K}}_{{\rm el},j}}{E^{300\,{\rm K}}_{{\rm el},j}}. \quad (8)$$

A single effective phonon temperature $\Theta$ is defined for all three poles, limiting correlations between parameters and providing consistency with other temperature-dependent components of the dispersion model. Multiple effective phonons were not necessary for the temperature range of the experimental data.

The spectral dependence of the refractive index determined by the dispersion model (1) is shown in Fig. 2(a) for several selected temperatures.

The parameters of the electron excitation dispersion model are listed in Table IV, and the temperature dependence of the electron excitation energies is shown in Fig. 3. While the excitation transition strengths $N_{{\rm el},j}$ are assumed to be temperature independent (reflecting the assumption of a constant number of valence electrons in the system), the corresponding contribution to the $f$-sum rule (2) decreases with temperature due to multiplication by the volume expansion factor $\mathcal{E}(T)$ in Eq. (1),

$$\int_0^\infty \mathcal{E}(T)\,E\,\chi_{{\rm i,el}}(E,T)\,{\rm d}E = \mathcal{E}(T)\sum_{j=1}^{3}N_{{\rm i},j}. \quad (9)$$

**TABLE IV.** Dispersion parameters describing electron valence-to-conduction band transitions.

| $j$ | $N_{{\rm el},j}$ (eV$^2$) | $E^{300\,{\rm K}}_{{\rm el},j}$ (eV) | $\varepsilon^{0\,{\rm K}}_{{\rm el},j}$ |
|---|---|---|---|
| 1 | 0.1458(16) | 1.4710(6) eV | 1.0616(3) |
| 2 | 59.4(2) | 2.749(3) eV | 1.0272(2) |
| 3 | 1652(8) | 14.76(2) eV | 0.9983(3) |
| $\Theta$ = 286(2) K | | | |

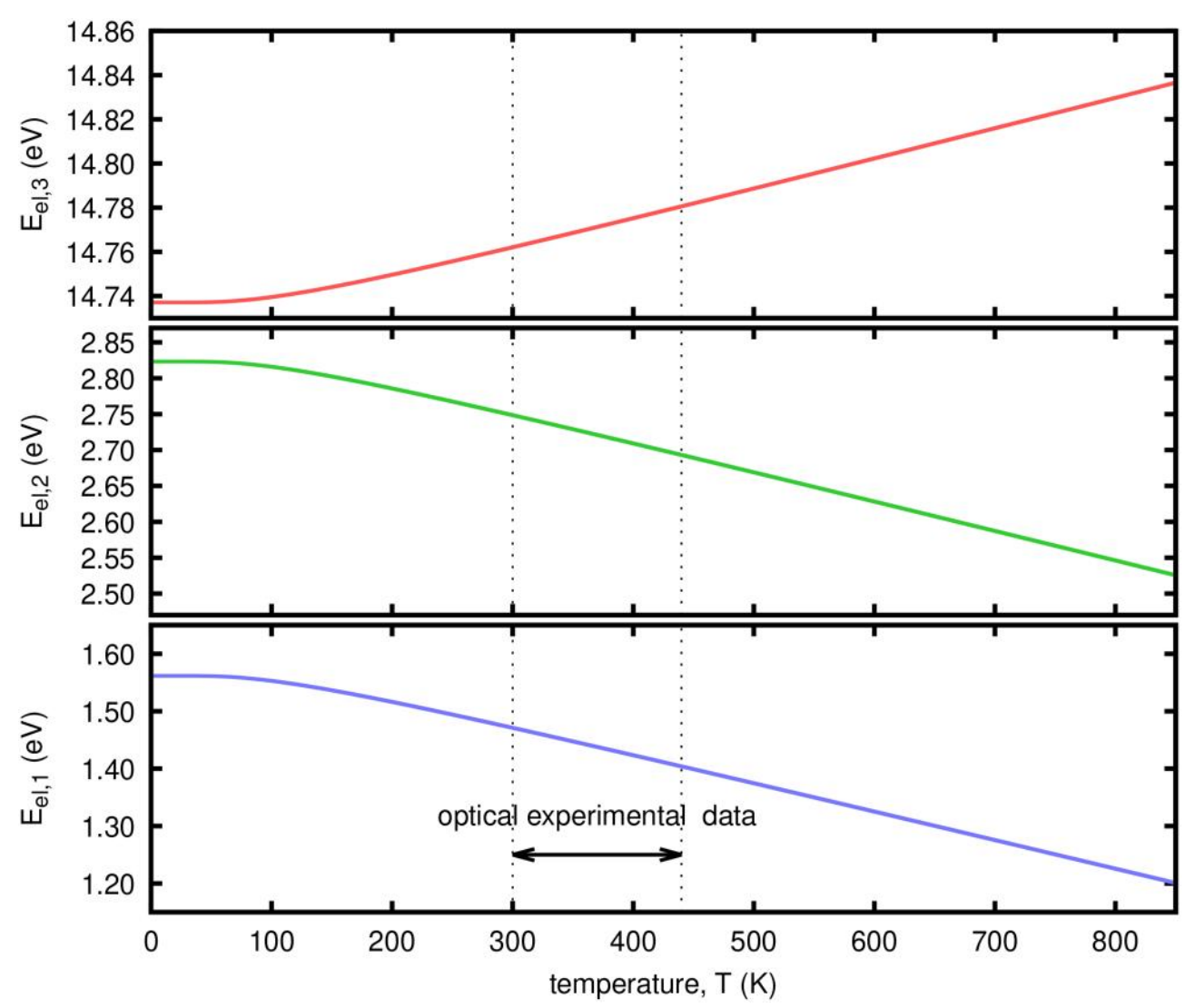

**FIG. 3.** Temperature dependence of electron excitation energies modeled by Bose–Einstein statistical factor (7) in harmonic oscillator model (6).

In addition, the effective strength of interband electronic excitations is expected to decrease with increasing temperature due to (i) thermal excitation of valence electrons into the conduction band and (ii) a redistribution of transition strength in favor of multi-phonon excitations.[7,27] Both effects would lead to a further reduction of the electronic contribution but are not included in the present dispersion model. The refractive index in the transparent region is, therefore, expected to decrease due to both thermal expansion and these neglected effects.

However, as seen in Fig. 2(a), the refractive index increases with temperature because spectral redistribution effects, i.e., the red shift of the electronic excitation bands, dominate over the reduction caused by thermal expansion.

A pronounced red shift is observed for the first two electronic excitation energies $E_{{\rm el},1}$ and $E_{{\rm el},2}$, whereas a weak blue shift is found for the highest excitation energy $E_{{\rm el},3}$ (see the parameters $\varepsilon^{0\,{\rm K}}_{{\rm el},j}$ in Table IV and Fig. 3). This blue shift of the highest excitation energy is most likely an artifact of the effective model, in which a relatively complex electronic absorption band is approximated by only three harmonic oscillators. The third Sellmeier term, therefore, acts as a compensating contribution that absorbs residual temperature-dependent effects not captured by the lower-energy oscillators.

Figure 3 shows that the excitation energies vary nearly linearly in the 300–440 K range. For a wider temperature range, the dependence could be extended by introducing multiple effective phonons, similar to the volume expansion factor. In an advanced model constructed in the future, common parameters $\Theta_1$, $\Theta_2$, and $\Theta_3$ of effective phonons could potentially be used for both electronic and phonon dispersion parameters.

In the study of Zn-doped samples, we assumed that the electronic excitation component is independent of dopant concentration, i.e., the distribution and density of valence electrons do not change with Zn doping. This approximation is valid because the dopant concentration is negligible compared to the Ga and As atomic density[6] ($4.4279 \times 10^{22}\,\mathrm{cm}^{-3}$). However, this does not imply that the refractive index in this region is completely independent of the dopants; Fig. 2(a) corresponds to undoped GaAs (sample #1).

Spectroscopic methods such as ellipsometry or spectrophotometry are not optimal for accurately determining the refractive index in transparent regions due to systematic measurement errors and non-ideal interfaces. The relative accuracy of these measurements is around $10^{-2}$. However, the measurement of fine features in the spectrum has a much higher reliability. Therefore, these non-idealities of the experimental data are compensated for by corrections within the model. These errors lead to a strong correlation between the thickness and the refractive index, especially when only far-IR spectrophotometric data with interference effects are available.

For precise refractive index determination, prism measurements using the minimum deviation method (MDM) are preferred. This method can achieve temperature-dependent refractive index accuracy in the range $1 \times 10^{-4}$– $1 \times 10^{-5}$, as demonstrated with the CHARMS equipment at NASA.[36] Lacking such equipment, we rely on previously published data, though for GaAs, highly precise data are limited. We found older MDM data from Marple[16] at 103, 187, and 300 K (0.73–1.39 eV) and room-temperature data from Hambleton *et al.*[17] (0.25–0.65 eV). For Hambleton *et al.*[17] data, we assumed a temperature of 20 °C, with deviations from 25 °C producing negligible differences ($\sim 10^{-3}$).

Skauli *et al.*[18] provided additional near-IR (0.6–1.28 eV) room temperature refractive index data derived from SHG measurements and transmittance interference patterns (0.073–0.83 eV, from room temperature to 95 °C). They modeled the data using a three-term Sellmeier formula and a two-term Pikhtin formula with one harmonic oscillator term in the far-IR. Temperature dependence was introduced empirically with some parameters linearly or quadratically dependent on $T$. The absolute refractive index accuracy from these methods is estimated at 0.2% ($7 \times 10^{-3}$), with SHG noise $\sim 2 \times 10^{-4}$.

Thus, the Skauli *et al.*[18] SHG data at $21 \pm 1$ °C complement the MDM data measured at room temperature and were used, together with the other datasets listed in Table II, to determine the parameters of our dispersion model (1). The room-temperature Skauli data exhibit the highest reliability and, therefore, serve as the primary reference for determining the absolute value of the GaAs refractive index in the transparent region, while all other datasets were fitted simultaneously. The consistency between these data and the model, as well as their differences from the model, is illustrated in Fig. 2.

Figure 4 shows the spectral uncertainties of the GaAs optical constants as a function of temperature and doping. The statistical uncertainties obtained from the regression method are much smaller than the expected uncertainty of the fitted refractive indices in the transparent region, which is estimated to be of the order of $\sim 10^{-3}$. This estimate is based on the declared uncertainties of the tabulated refractive index data used in the fitting procedure, as discussed above. Furthermore, the fitting procedure was applied not only to the tabulated refractive index data, but also to all experimental and tabulated data listed in Table II. As a result, the obtained optical constants include phonon absorption features, which will be discussed in detail in Secs. IV C–IV D.

As shown in Fig. 2(b), the refractive index differences calculated using the Sellmeier and Pikhtin formulas from Skauli *et al.*[18] and the dispersion model (1) are within $\pm 1 \times 10^{-3}$ up to 1.28 eV at 300 and 370 K. At higher energies and 440 K, deviations increase due to lack of experimental FTIR and SHG data in these spectral and temperature ranges. Our results extrapolate beyond 300 K in the mid- and near-IR regions, producing a relatively good agreement but with systematic deviations at 300, 370, and 440 K.

The temperature dependence of the dispersion model (1) was determined from the shift of interference patterns observed in high-resolution far-infrared transmittance and reflectance measurements (Fig. 5). Accurate modeling of both single-phonon and multi-phonon absorption is, therefore, required to correctly interpret these interference patterns, as discussed in Secs. IV C–IV D.

The average thickness of the GaAs wafer was determined as $d_{s,1} = 0.628\,061(14)\,\mathrm{mm}$ with a thickness variation of $\sigma(d_{s,1}) = 0.001\,144(5)$ mm. The spectral resolution of the measured data was fitted, yielding a full width at half maximum of $1.1051(11)\,\mathrm{cm}^{-1}$, in good agreement with the nominal spectrophotometer resolution of $1\,\mathrm{cm}^{-1}$. Thickness non-uniformity and spectral resolution are expected to be anti-correlated, as both effects lead to attenuation of the interference fringes. However, the corresponding correlation coefficient obtained from the fit is $-0.2501$, indicating only a weak correlation. This explains the relatively small uncertainties of the fitted parameters.

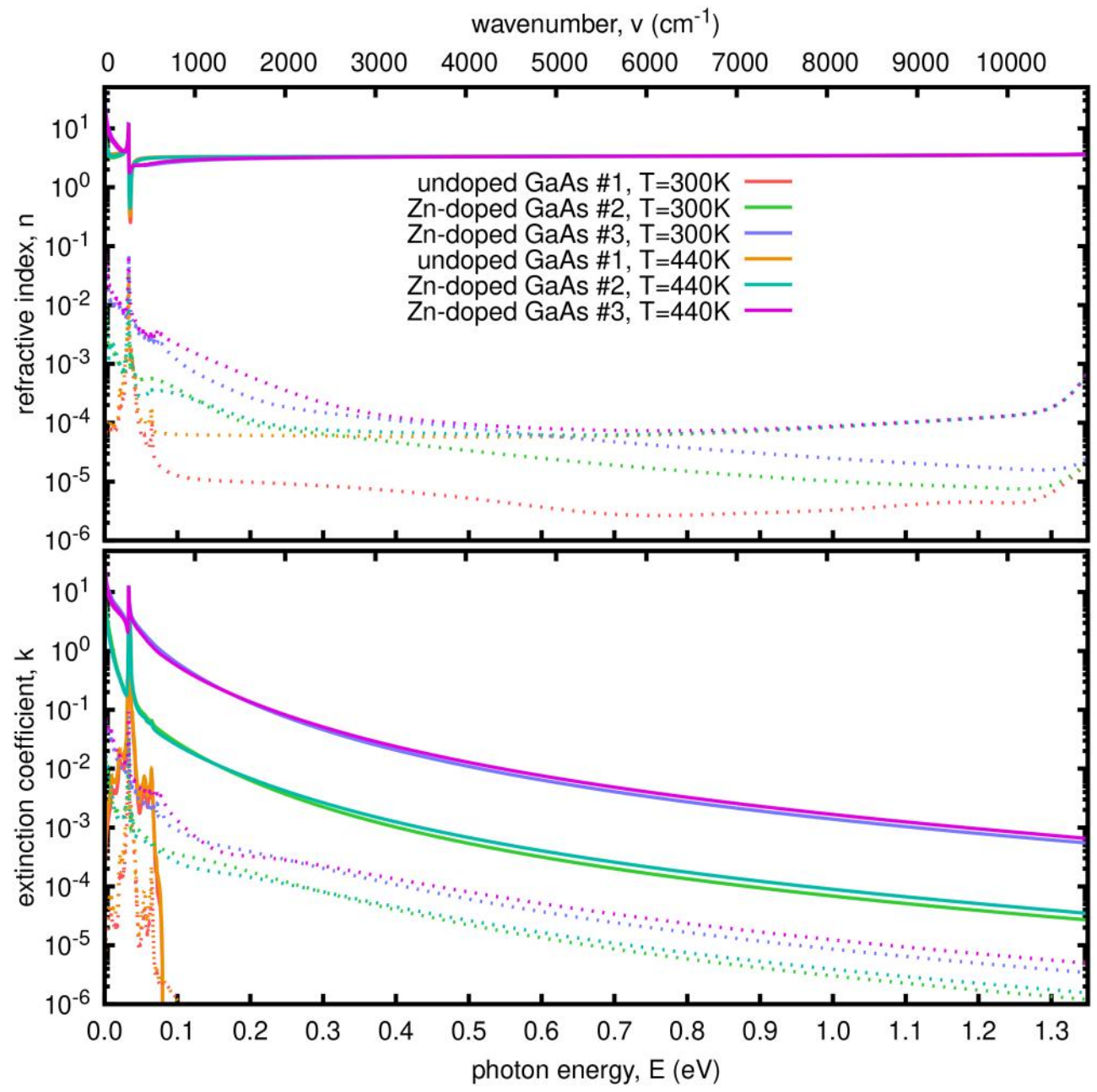

**FIG. 4.** Spectral dependencies of the refractive index $n(E)$ and extinction coefficient $k(E)$ of GaAs samples (full lines) and their uncertainties (dotted lines) determined from the regression method used.

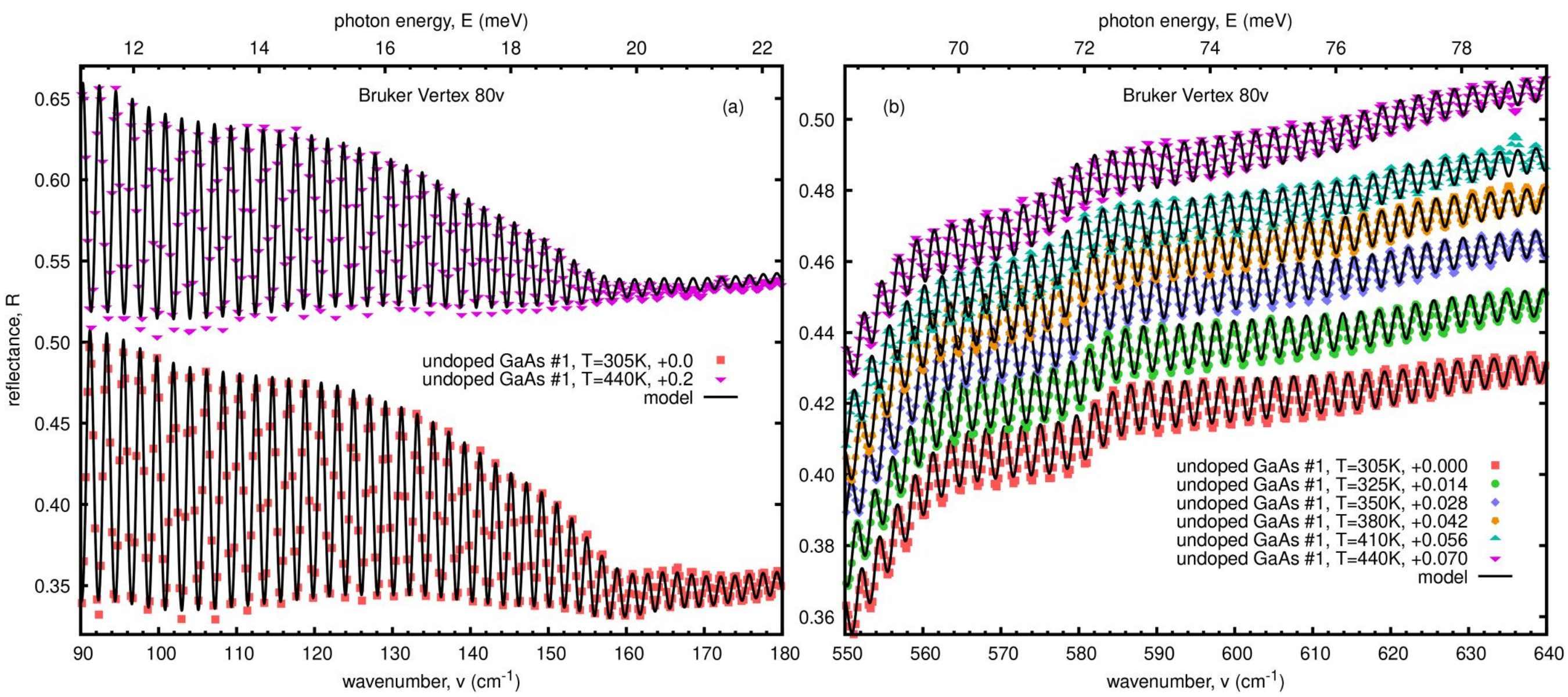

**FIG. 5.** High-resolution reflectance spectra of undoped GaAs measured at selected temperatures in (a) the partially transparent region below the TO-phonon resonance frequency and (b) above the LO-phonon resonance frequency. Spectra are offset as indicated in the legend.

Similarly, Skauli *et al.*[18] determined the temperature dependence of the refractive index from interference fringe shifts observed in their mid-infrared transmittance measurements. Figure 2(b), therefore, shows small but systematic deviations between the refractive index obtained from the present dispersion model and the values reported by Skauli *et al.*[18] in the overlap energy range (0.075–0.085 eV). These deviations originate from the assumption of a linear thermal expansion coefficient adopted in Ref. 18, which introduces a systematic error into the temperature dependence of the refractive index [see Eq. (8) in Ref. 18].

Figure 1 compares the linear expansion model used by Skauli *et al.*[18] with the temperature-dependent expansion coefficient given by Eq. (5), which is incorporated into the present dispersion model (1) and used to describe the temperature dependence of the substrate thickness via Eq. (4). This comparison demonstrates that the choice of the thermal expansion model is the primary source of the systematic deviations observed in Fig. 2(b).

The aim of this section is not to determine highly accurate temperature dependencies of the refractive index in the mid- and near-infrared regions, although the dispersion model yields reasonable agreement there. Rather, the primary goal is to obtain a reliable effective description of the electronic contributions in the far-infrared region, i.e., in the phonon excitation range, which is essential for modeling the infrared response discussed in the remainder of this work.

### C. Single-phonon excitations

Cubic GaAs crystallizes in the zinc blende structure, which exhibits one dielectrically active TO phonon[6] at 8.02 THz. It is known that the isotopic composition of individual atoms affects the frequency and distribution of phonons.[37] In diatomic crystals, such as GaAs, the TO-phonon frequency is inversely proportional to the reduced mass of the system of individual atoms.[37] While arsenic occurs in nature as one stable isotope, $^{75}$As, gallium occurs in two stable isotopes, i.e., $^{69}$Ga (60.1%) and $^{71}$Ga (39.9%). Simple algebra shows that, if we were to study isotopically pure crystals, a crystal with light isotopes would have a TO-phonon frequency 1.5% higher than a crystal with heavy isotopes. The TO-phonon frequencies of natural crystals lie somewhere between these values.

Furthermore, isotopically pure crystals have a narrower frequency distribution than crystals with randomly distributed isotopes. Moreover, if the occurrence of individual isotopes is unequal, it can be assumed that this may partially explain the lack of complete symmetry in the absorption peak.

Traditionally, the reflectivity in the reststrahlen region and its vicinity has been used to study single-phonon excitations.[38] The use of ellipsometry, which is sensitive to surface effects, is also possible.[39] Classical dispersion models, i.e., the Drude–Lorentz model, have been successfully used to interpret experimental data.[38,39]

Figure 6 shows fits of reflectivity measured at high resolution ($1\,cm^{-1}$) using three discrete excitation models with different distribution functions, i.e., Lorentz, Gaussian, and Voigt. The fitted experimental data were limited to the spectral range 230–320 $\mathrm{cm}^{-1}$, where sample #1 was non-transparent, i.e., free of interference patterns. These fits are for demonstration purposes only and are not used further in this work. The exact parameterization of dispersion models are given in Appendix A.

Table V lists the corresponding dispersion parameters and the $\chi$ values describing the quality of the fit. Figure 6 also shows the final fit of the optical characterization, based on processing the complete set of experimental data. To compare the fits, we calculated the partial $\chi$ value corresponding to the fit of the final model to the reflectance data in Fig. 6, which was 1.546.

As seen, the classical model, i.e., a discrete excitation with a Lorentzian line shape, describes the experimental data quite well. In contrast, the use of a Gaussian profile results in a poor fit. The absorption peak becomes broader and spectrally shifted, and the corresponding loss function is strongly suppressed. In fact, the loss function exhibits a small maximum in the vicinity of the TO frequency, followed by a very weak local maximum near the LO frequency; see Fig. 6. Thus, the Gaussian model fails to reproduce the experimental response.

If a Voigt profile is used, the fit improves only slightly compared to the classical model. The absorption peak is somewhat lower and broader, but the peak in the loss function remains practically the same as in the classical model.

The best fit is obtained using the final model, which will be discussed below. The absorption peak is slightly lower and wider compared to the classical and Voigt models, but the peak in the loss function remains nearly identical to the classical model.

Note that the Lorentzian discrete excitation dispersion model used here is equivalent to the well-known Lorentzian model. It differs only in parameterization, using the resonant frequency instead of the central frequency, which are practically identical for small broadening parameters. Additionally, our models normalize excitation strengths using the f-sum rule so that individual excitations can be correctly used in a temperature-dependent model. Further details can be found in previous articles[24,40] or in Appendix A.

In addition, Fig. 6 shows the positions of the excitation energies of TO and LO phonons, which in the classical model were directly model parameters when using the factorized form of the Lorentz model (see Appendix A). The frequencies obtained from the final model will be discussed later. The TO-phonon excitation energies are given in Table V. The LO-phonon excitation energy $E_{\mathrm{LO}}$ was 36.1368(6) meV (291.5 $\mathrm{cm}^{-1}$, 8.738 THz).

From the broadening parameters, it is evident that the Lorentzian component plays a more significant role in the Voigt profile than the Gaussian component. Thus, it was logical to base the final model on distributions containing the Lorentzian component.

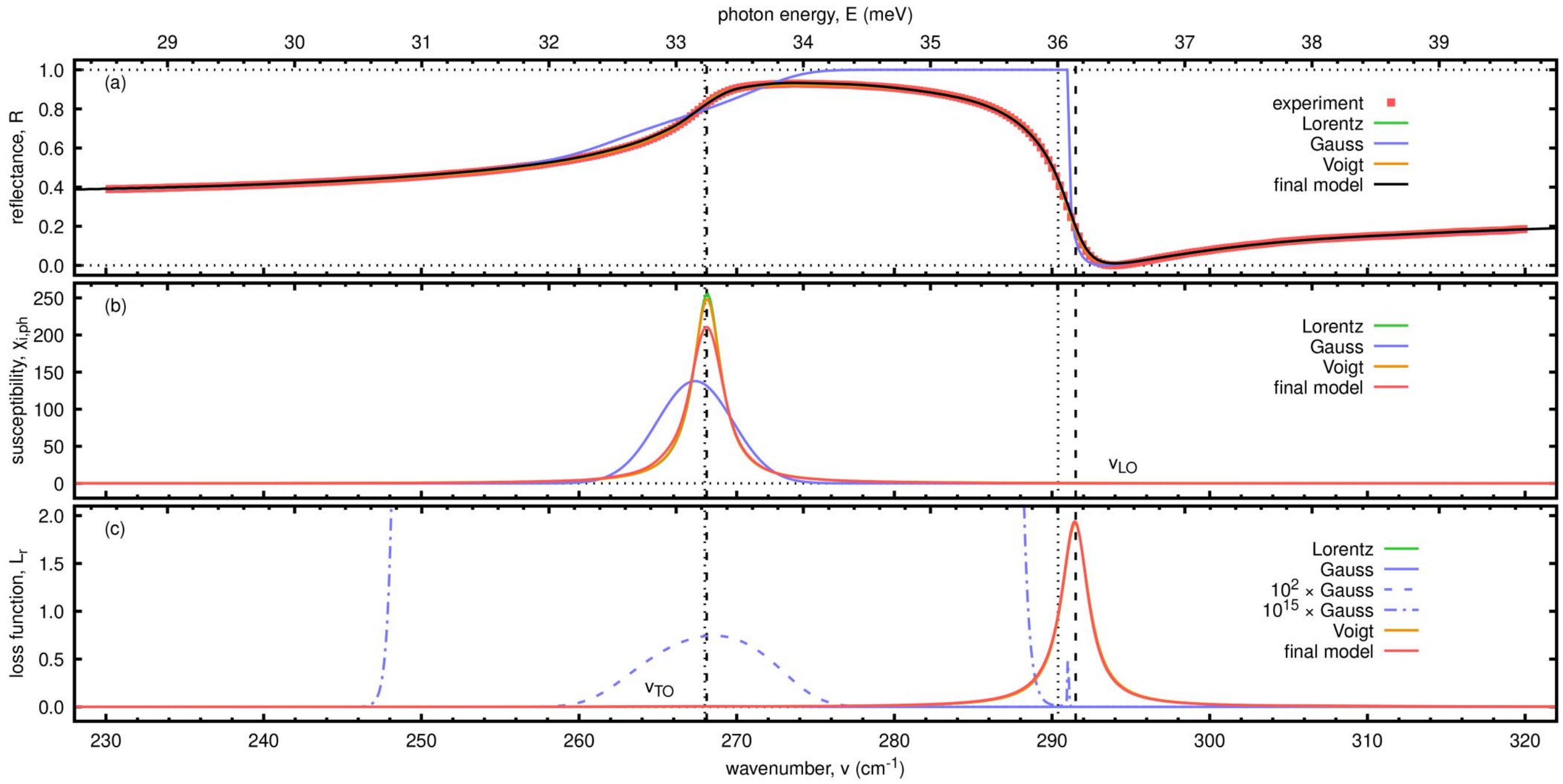

**FIG. 6.** Reflectance and corresponding spectral dependence of the response function of single-phonon excitations of undoped GaAs determined using several dispersion models in the reststrahlen region: (a) high-resolution reflectance data; (b) imaginary part of susceptibility $\chi_{\mathrm{i,ph}}(E)$; (c) loss function $L_{\mathrm{r}}(E)$, where $L(E) = -\mathrm{i}\hat{\varepsilon}^{-1}(E)$. The vertical lines indicate the TO- and LO-phonon resonance frequencies obtained from the final model (dotted lines) and from the classical model (dashed lines). The lines for the TO-phonon frequencies are so close that they merge in this image (for details, see Fig. 8).

**TABLE V.** Dispersion parameters describing single-phonon excitations: $N$—transition strength; $E_{TO}$—TO-phonon excitation energy; Γ—Lorentzian broadening parameter (FWHM); $B$—Gaussian broadening parameter (RMS value); $\varepsilon_\infty$—contribution of electron excitations. Parameter $\chi$ describes the quality of the fit (proportional to the mean square value of differences).

| Parameter | Lorentz | Gaussian | Voigt |
|---|---|---|---|
| $N$ ($10^{-3}$eV$^2$) | 3.485(4) | 3.40(13) | 3.484(4) |
| $E_{TO}$ (meV) | 33.2439(12) | 33.148(9) | 33.2430(13) |
| Γ (meV) | 0.2613(7) | … | 0.2609(7) |
| $B$ (meV) | … | 0.30(3) | 0.022(6) |
| $\varepsilon_\infty$ | 11.055(12) | 10.8(4) | 11.050(13) |
| $\chi$ | 2.724 | 103.7 | 2.71 |

To interpret the complete set of experimental data (see Fig. 7), we tested numerous distribution functions. It is important to note that the classical Lorentzian model was completely inadequate, because the Lorentzian function decreases too slowly away from the resonant frequency, making it impossible to interpret the transmittance data further from the reststrahlen region, where sample #1 was transparent (see Fig. 5).

Due to the different masses of gallium isotopes, we attempted a binomial distribution of resonance frequencies to explain the weak asymmetry of the absorption peak observed in reflectance, but without success. Gradually, we increased the number of excitation energies within the asymmetric Voigt distribution model. As the number of absorption peaks increased, the Lorentzian components of the Voigt profiles surprisingly decreased. When fitting included about five peaks, these components almost disappeared, and the experimental data were well fitted by five nearly Gaussian peaks.

In the final model, we used six Gaussian-broadened peaks with complex amplitudes within the asymmetric peak approximation[24,40]

$$\hat{\chi}_{\mathrm{ph}}(E,T)=\sum_{j=1}^{6}\left(N_{\mathrm{ph},j}+\mathrm{i}M_{\mathrm{ph},j}\frac{E_{\mathrm{ph},j}(T)}{E}\right)\hat{\chi}^{0}_{\mathrm{G},j}(E,T), \tag{10}$$

where $\hat{\chi}^{0}_{\mathrm{G},j}(E,T)$ is normalized Gaussian contribution calculated as

$$\hat{\chi}^{0}_{\mathrm{G},j}(E,T)=\frac{\mathrm{i}}{\sqrt{2\pi}B_{\mathrm{ph},j}(T)E_{\mathrm{ph},j}(T)}\times\left\{\frac{2\mathrm{i}}{\sqrt{\pi}}[\mathrm{D}(X_-)-\mathrm{D}(X_+)]+\left[\exp\left(-X_-^2\right)-\exp\left(-X_+^2\right)\right]\right\}, \tag{11}$$

with

$$X_\pm=\frac{E\pm E_{\mathrm{ph},j}(T)}{\sqrt{2}B_{\mathrm{ph},j}(T)}$$

and D(·) denotes the Dawson function[41,42] (A7). Normalization is performed according to the f-sum rule:

$$\int_0^\infty E\chi^{0}_{\mathrm{i,G},j}(E,T)\,\mathrm{d}E=1. \tag{12}$$

The parameters $N_{\mathrm{ph},j}$ and $M_{\mathrm{ph},j}$ determine the transition strengths and the coupling parameters, in other words, the excitation (harmonic) and interaction (anharmonic) parts of the response function. Note that the f-sum rule excluding volume expansion factor $\mathcal{E}(T)$ is only a function of the parameter $N_{\mathrm{ph},j}$ and does not depend on the parameters $M_{\mathrm{ph},j}$,

$$\int_0^\infty E\chi_{\mathrm{i,ph}}(E,T)\,\mathrm{d}E=\sum_j N_{\mathrm{ph},j}=N_{\mathrm{ph}}. \tag{13}$$

Fit of IR ellipsometry data contributes little to the exact shape of the response function in this region as it lies at the instrument's limit. It was used mainly to determine the native GaAs layer thickness, which slightly affects reflected and transmitted IR light. The optical constants of the native layer were assumed equal to those of amorphous gallium oxide.[43] The layer was assumed identical for all three samples, with thickness 3.31(2) nm.

The absorption peak shape is, thus, defined by an exact fit of the reflectance data in and around the reststrahlen region (see Fig. 7).

The dispersion parameters are summarized in Table VI. The internal parameter $M^{\#1}_{\mathrm{ph},1}$ is calculated from the condition

$$\hat{\chi}_{\mathrm{ph}}(0,T)=0, \tag{14}$$

which ensures that sample #1 (undoped GaAs) is described by the response function of a non-conducting material. The peak asymmetry parameters $M^{\#1}_{\mathrm{ph},j}$ describe the coupling of individual vibrational modes representing the TO phonon in a crystal with randomly distributed isotope masses. As shown in Ref. 24, such coupling manifests as mutual phase shifts of vibrational modes compared to an uncoupled system.

From the uncertainties of the $N_{\mathrm{ph},j}$ and $M_{\mathrm{ph},j}$ parameters, it is evident that the model is somewhat overparameterized. However, this does not affect the obtained optical constants of GaAs (see Fig. 4). In an advanced model, $M_{\mathrm{ph},j}$ will depend on the dopant concentration. Here, we assume a linear variation of $M_{\mathrm{ph},j}$ with Zn concentration via the parameter $p^{\#2}_{\mathrm{ph}}$,

$$M^{\#2}_{\mathrm{ph},j}=\left(1-p^{\#2}_{\mathrm{ph}}\right)M^{\#1}_{\mathrm{ph},j}+p^{\#2}_{\mathrm{ph}}M^{\#3}_{\mathrm{ph},j}. \tag{15}$$

Similarly, for the response functions of individual samples,

$$\hat{\chi}^{\#2}_{\mathrm{ph}}(E,T)=\left(1-p^{\#2}_{\mathrm{ph}}\right)\hat{\chi}^{\#1}_{\mathrm{ph}}(E,T)+p^{\#2}_{\mathrm{ph}}\hat{\chi}^{\#3}_{\mathrm{ph}}(E,T). \tag{16}$$

The assumed linear dependence follows from the fact that the interaction terms depend on the densities and distribution functions of the coupled subsystems. If the phonon excitation spectrum itself is assumed to be independent of the dopant concentration, a linear dependence of the interaction terms on the dopant concentration naturally arises, provided that the free-carrier distribution function does not change significantly with doping. As will be shown in Sec. IV E, this assumption is not strictly fulfilled. Nevertheless, a monotonic dependence of the interaction terms on the dopant concentration can be expected, and the linear dependence, therefore, represents a reasonable first-order approximation.

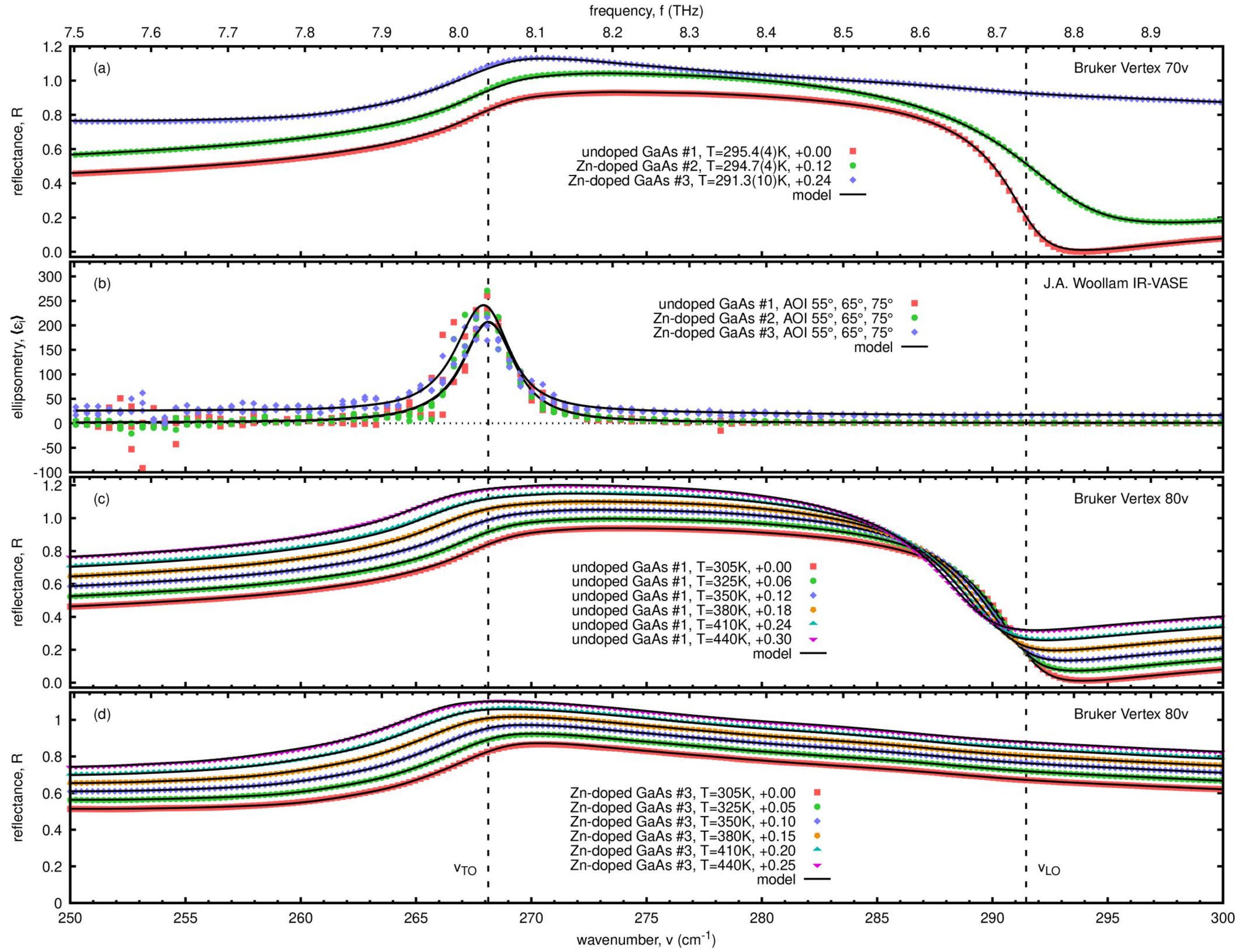

**FIG. 7.** Selected experimental data in the region of single-phonon excitations in GaAs: (a) reflectance measurement at room temperature; (b) ellipsometric measurement at room temperature (imaginary part of pseudodielectric function); (c) temperature-dependent reflectance measurement of sample #1; (d) temperature-dependent reflectance measurement of sample #3. Reflectance data are offset as indicated in the legend. The vertical dashed line indicates the reststrahlen region, bounded by the TO-phonon frequency 8.0384 THz (268.13 $cm^{-1}$, 33.244 meV) and LO-phonon frequency 8.7378 THz (291.46 $cm^{-1}$, 36.137 meV) determined in this work using the factorized Lorentz model.

From $p^{\#2}_{\mathrm{ph}}$, it can be inferred that the doping concentration of sample #2 is 5.7% of that of sample #3, assuming a linear relation between coupling and free-carrier density.

The internal parameter $M^{\#1}_{\mathrm{ph},1}$ is weakly temperature dependent, unlike the other $M_{\mathrm{ph},j}$ parameters. From a theoretical standpoint, there is no reason for $M_{\mathrm{ph},j}$ to be temperature independent, unlike $N_{\mathrm{ph},j}$ whose temperature independence results from sum rules. For wider temperature ranges, $M_{\mathrm{ph},j}$ may need to be considered temperature dependent.

Figure 8(a) shows the distribution of discrete excitations and broadening in the final model. The resulting absorption peak is slightly asymmetric, as shown in Fig. 8(b). For comparison, symmetric Lorentz and Voigt peaks are also shown. This asymmetry can be explained by the uneven distribution of Ga isotopes; however, it cannot be ruled out that coupling of TO phonons with multi-phonon excitations also plays a role.

The temperature dependence of the excitation energies and broadening parameters is introduced analogously to electron excitations,

**TABLE VI.** Dispersion parameters describing single-phonon excitations: $N_{\mathrm{ph},j}$—transition strength; $E^{300\,\mathrm{K}}_{\mathrm{ph},j}$—excitation energy; $B^{300\,\mathrm{K}}_{\mathrm{ph},j}$—broadening parameter (RMS); $M^{\#1}_{\mathrm{ph},j}$, $M^{\#3}_{\mathrm{ph},j}$—parameters determining coupling (peak asymmetry); $\varepsilon^{0\,\mathrm{K}}_{\mathrm{ph}}$—ratio of excitation energies at $T = 0$ K and RT; $\beta^{0\,\mathrm{K}}_{\mathrm{ph}}$—ratio of broadening at $T = 0$ K and RT; Θ—effective phonon temperature; and $p^{\#2}_{\mathrm{ph}}$—parameter determining $M^{\#2}_{\mathrm{ph},j}$.

| $j$ | $N_{\mathrm{ph},j}$ ($10^{-6}$ eV$^2$) | $E^{300\,\mathrm{K}}_{\mathrm{ph},j}$ (cm$^{-1}$) | $B^{300\,\mathrm{K}}_{\mathrm{ph},j}$ (cm$^{-1}$) | $M^{\#1}_{\mathrm{ph},j}$ ($10^{-6}$ eV$^2$) | $M^{\#3}_{\mathrm{ph},j}$ ($10^{-6}$ eV$^2$) |
|---|---|---|---|---|---|
| 1 | 266(230) | 270.3(11) | 4.2(6) | Eq. (14) | 99(254) |
| 2 | 934(47) | 267.99(4) | 0.804(13) | −272(84) | 146(97) |
| 3 | 204(151) | 274(3) | 5.6(10) | −161(216) | −33(204) |
| 4 | 375(219) | 266(3) | 7.0(8) | 105(121) | 549(399) |
| 5 | 298(229) | 266.8(2) | 1.49(10) | −802(171) | −1154(224) |
| 6 | 1207(217) | 267.3(3) | 2.28(7) | 737(106) | 1221(229) |

$\varepsilon^{0\,\mathrm{K}}_{\mathrm{ph}} = 1.012\,03(7)$, $\beta^{0\,\mathrm{K}}_{\mathrm{ph}} = 0.890(8)$, $\Theta = 286(2)$ K, $p^{\#2}_{\mathrm{ph}} = 0.057(7)$

$$\begin{aligned} E_{\mathrm{ph},j}(T) &= E^{300\,\mathrm{K}}_{\mathrm{ph},j}\left[\varepsilon^{0\,\mathrm{K}}_{\mathrm{ph}} + \left(1-\varepsilon^{0\,\mathrm{K}}_{\mathrm{ph}}\right)\frac{\exp(\Theta/T_{\mathrm{R}})-1}{\exp(\Theta/T)-1}\right],\\ B_{\mathrm{ph},j}(T) &= B^{300\,\mathrm{K}}_{\mathrm{ph},j}\left[\beta^{0\,\mathrm{K}}_{\mathrm{ph}} + \left(1-\beta^{0\,\mathrm{K}}_{\mathrm{ph}}\right)\frac{\exp(\Theta/T_{\mathrm{R}})-1}{\exp(\Theta/T)-1}\right],\\ \varepsilon^{0\,\mathrm{K}}_{\mathrm{ph}} &= \frac{E^{0\,\mathrm{K}}_{\mathrm{ph},j}}{E^{300\,\mathrm{K}}_{\mathrm{ph},j}}, \quad \beta^{0\,\mathrm{K}}_{\mathrm{ph}} = \frac{B^{0\,\mathrm{K}}_{\mathrm{ph},j}}{B^{300\,\mathrm{K}}_{\mathrm{ph},j}}. \end{aligned} \tag{17}$$

To reduce the number of dispersion parameters, the effective phonon temperature Θ in the Bose–Einstein statistics could be equal to that of the electron excitation.

Figure 9 shows that excitation energies decrease and broadening increases with temperature, nearly linearly from 300 to 440 K.

From Fig. 10(a), it is evident that free carriers have a relatively small effect on the lightly doped GaAs sample #2, whereas their effect on the heavily doped GaAs sample #3 is more pronounced. For sample #3, a detailed comparison reveals a subtle asymmetry of the absorption peak, although at first glance, the effect may appear as a simple red shift of the peak position. This apparent shift is in fact caused by peak asymmetry, while all three absorption peaks correspond to the same excitation energies (resonance frequencies). In other words, coupling with free carriers does not change the resonance frequency, but alters the phase with which the vibrational states respond to the mean field, which manifests as asymmetry in the frequency-domain absorption peak. As known from the theory, coupling does not affect the absorption peak corresponding to free carriers.[24] When subtracting the TO-phonon response function of the system with free carriers from that of the system without free carriers, an interaction term describing the coupling between the TO phonons and the free carriers is obtained,

$$\hat{\chi}^{\#s}_{\mathrm{C(ph,fc)}}(E,T) = \hat{\chi}^{\#s}_{\mathrm{ph}}(E,T) - \hat{\chi}^{\#1}_{\mathrm{ph}}(E,T). \tag{18}$$

This interaction part of the response function is shown in Fig. 10(b).

It should be noted that the reflectance spectra of all three samples can also be fitted reasonably well using a conventional Drude–Lorentz model without explicit phonon–free-carrier coupling. In such a description, the TO-phonon absorption peak of the heavily doped sample #3 appears shifted to lower frequencies by approximately 1% compared to the undoped sample. Considering that only about $10^{-4}$ of the lattice atoms are substituted by Zn, such a large apparent shift of the phonon resonance frequency is difficult to justify in terms of a genuine modification of the lattice eigenfrequencies. This indicates that the apparent phonon frequency shift obtained within an uncoupled model is an effective artifact, arising from the neglect of phonon–free-carrier coupling, rather than a true change of the phonon resonance energy.

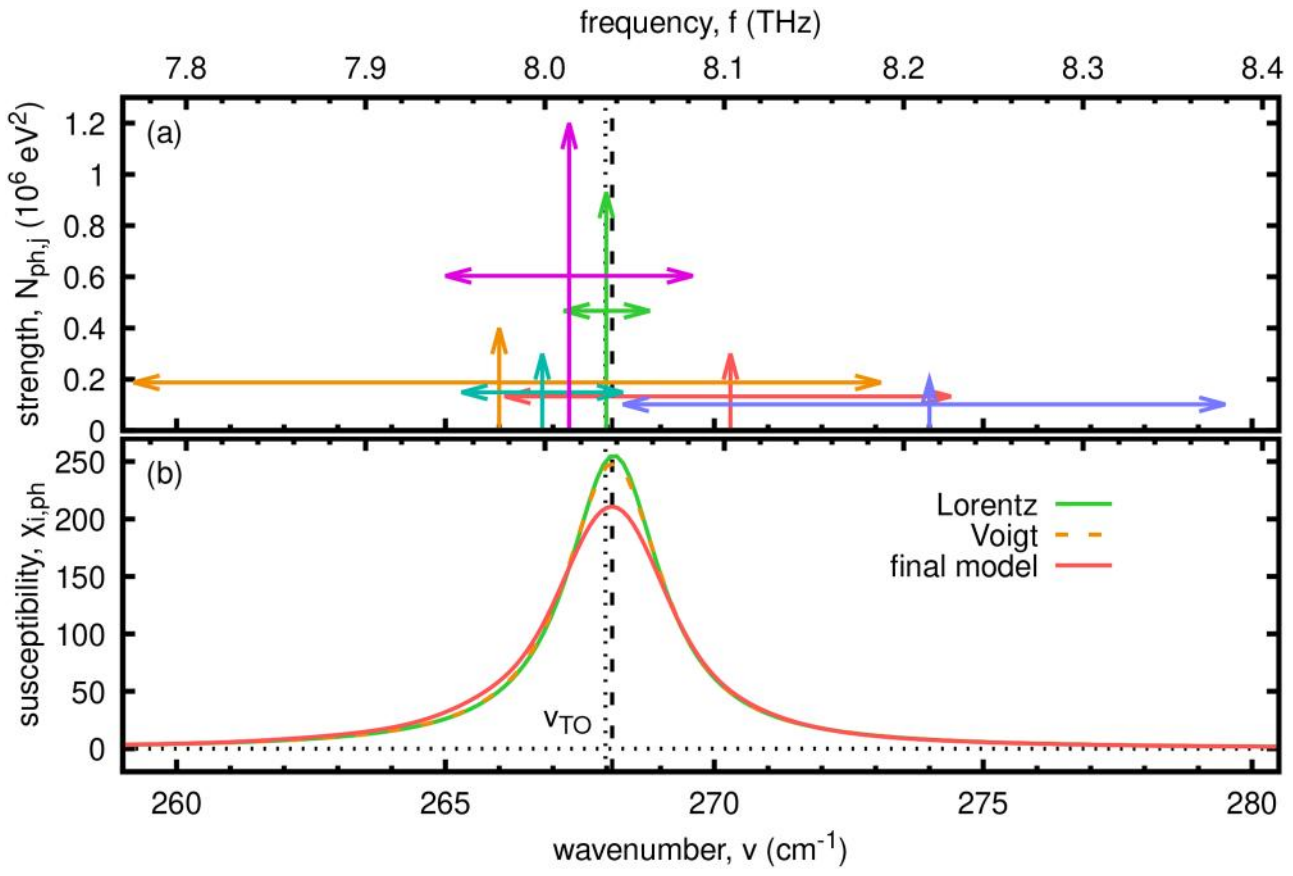

**FIG. 8.** Single-phonon excitations for undoped GaAs at RT. (a) Vertical arrows indicate the positions of the six individual resonance frequencies modeling TO phonons. Horizontal arrows with the same color show corresponding broadening (length $2B^{300\,\mathrm{K}}_{\mathrm{ph},j}$). Color coding is consistent with Fig. 9. (b) The imaginary part of susceptibility $\chi_{i,\mathrm{ph}}(E)$ calculated by three models. Vertical dashed lines indicate TO-phonon resonance frequencies as in Fig. 6.

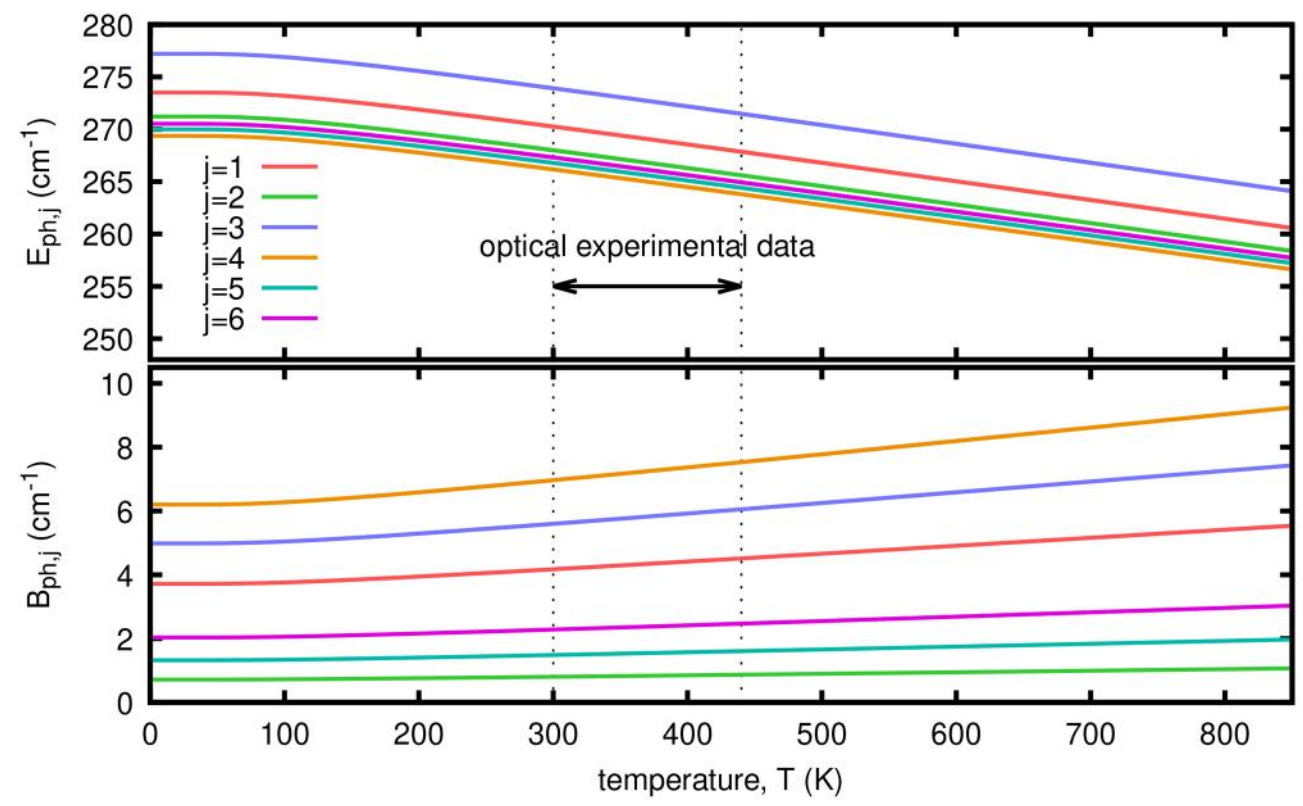

**FIG. 9.** Temperature dependence of phonon excitation energies and broadening parameters.

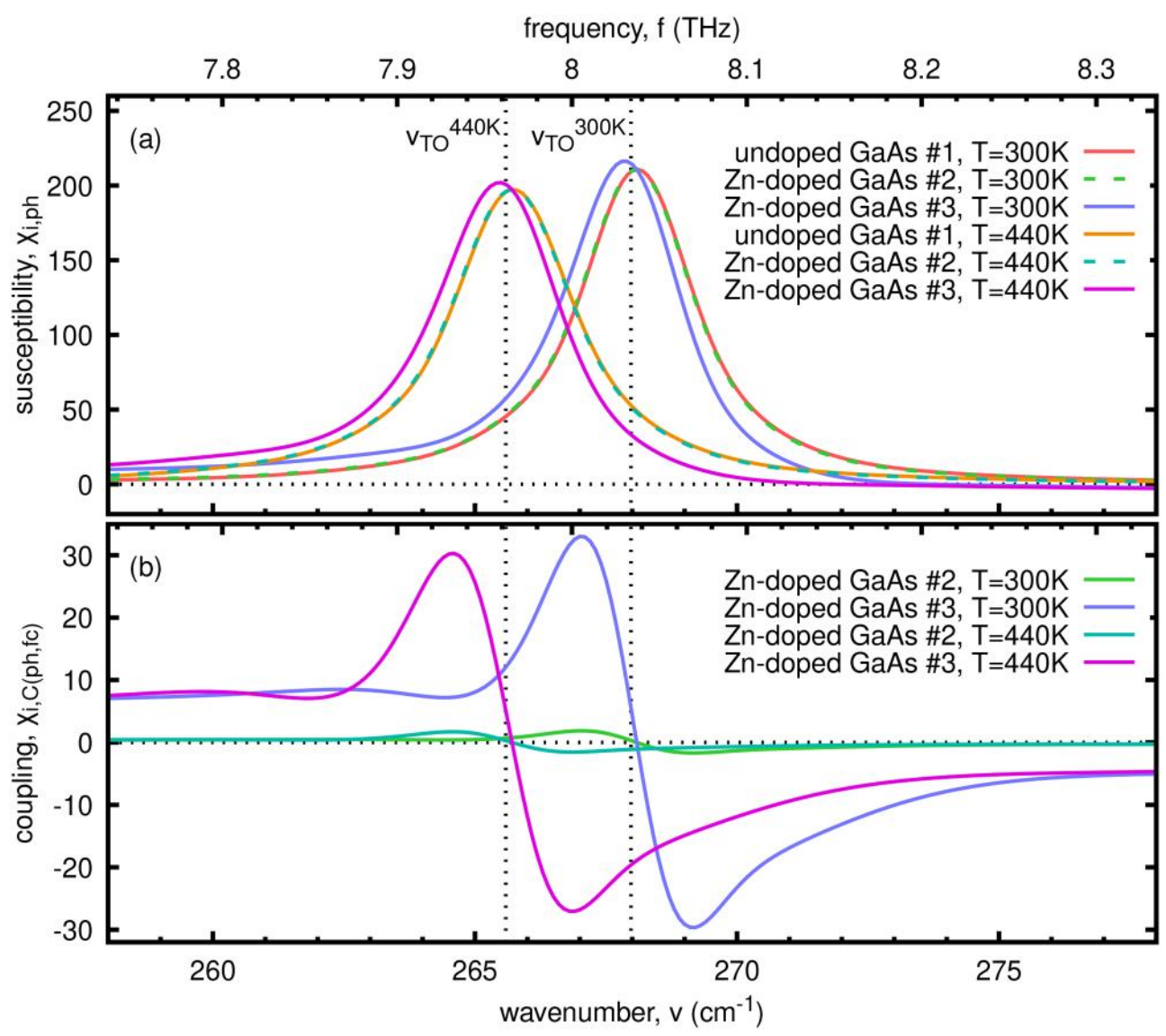

FIG. 10. (a) Spectral dependence of the imaginary part of the susceptibility of single-phonon excitations as a function of temperature and Zn concentration. Vertical dotted lines indicate the mean TO-phonon resonance frequency at 300 and 440 K, calculated as a weighted mean using (19). (a) Corresponding interaction part describing the coupling of TO phonons with free carriers calculated as (18).

The weighted mean energy of single-phonon excitations representing TO phonons for all three samples is

$$E_{\mathrm{TO}} = \frac{\sum_{j=1}^{6} N_{\mathrm{ph},j} E_{\mathrm{ph},j}}{\sum_{j=1}^{6} N_{\mathrm{ph},j}} = 33.225\,\mathrm{meV} \tag{19}$$

($267.98\,\mathrm{cm}^{-1}$, 8.0337 THz) for a temperature of 300 K. This value agrees well with the frequency obtained from the classical model applied to the reflectivity around the reststrahlen region [33.2439(12) meV, see Fig. 8].

Furthermore, although the peak for sample #3 may initially appear higher, the lattice polarizability of all three samples is the same because all peaks are described by the same total transition strength

$$N_{\mathrm{ph}} = \sum_j N_{\mathrm{ph},j} = 3.2848 \times 10^{-3}\,\mathrm{eV}^2. \tag{20}$$

In other words, the f-sum rule integral is the same for all three response functions describing single-phonon excitations, i.e., it is independent of the $M_{\mathrm{ph},j}$ parameters.

In the mid- and near-IR region, the parameters $E_{\mathrm{TO}}$ and $N_{\mathrm{ph}}$ can be used in a simple temperature-dependent harmonic oscillator (HO) model so that the phonon excitations are replaced by a harmonic oscillator. Together with the three harmonic oscillators describing the electron excitations (6) and the volume expansion factor (3) calculated via (5), this yields

$$n(E,T) = \left\{1 + \mathcal{E}(T)\frac{2}{\pi}\left[\frac{N_{\mathrm{ph}}}{E_{\mathrm{TO}}^2(T) - E^2} + \sum_{j=1}^{3}\frac{N_{\mathrm{el},j}}{E_{\mathrm{el},j}^2(T) - E^2}\right]\right\}^{1/2}. \tag{21}$$

The temperature dependence of $E_{\mathrm{TO}}$ is included via (17).

As mentioned earlier, the contribution of electron excitations in the phonon region can be approximated by the constant $\varepsilon_\infty - 1$. For the HO model at room temperature (RT), the following approximation holds:

$$\varepsilon(E, T_{\mathrm{R}}) \approx \varepsilon_\infty + \frac{2}{\pi}\frac{N_{\mathrm{ph}}}{E_{\mathrm{TO}}^2 - E^2} = \varepsilon_\infty \frac{E_{\mathrm{LO}}^2 - E^2}{E_{\mathrm{TO}}^2 - E^2},$$
$$\varepsilon_\infty = 1 + \frac{2}{\pi}\sum_{j=1}^{3}\frac{N_{\mathrm{el},j}}{E_{\mathrm{el},j}^{300\,\mathrm{K}\,2}} = 10.873, \tag{22}$$

where the term expressed in terms of the LO-phonon resonance energy $E_{\mathrm{LO}}$ corresponds to the factorized form of the HO model. Using this factorized model, $E_{\mathrm{LO}}$ can be calculated as

$$E_{\mathrm{LO}} = \sqrt{E_{\mathrm{TO}}^2 + \frac{2}{\pi}\frac{N_{\mathrm{ph}}}{\varepsilon_\infty}} = 36.003\,\mathrm{meV} \tag{23}$$

($290.38\,\mathrm{cm}^{-1}$, 8.7055 THz). The $E_{\mathrm{LO}}$ value does not match the position of the maximum in the loss function nor the value obtained from the classical model (see Fig. 6 and Table V). This underestimation occurs even though all quantities on the right-hand side of Eq. (23) are accurately determined. The final model also differs in the values of $N_{\mathrm{ph}}$ and $\varepsilon_\infty$ used in Eq. (23). One limitation is that the HO model cannot accurately determine $E_{\mathrm{LO}}$ if the peak distribution is non-Lorentzian. For non-Lorentzian distributions, it is easier to extract $E_{\mathrm{LO}}$ from the maximum of the loss function than to derive an exact expression. This issue will be discussed further in Sec. IV D.

Finally, we emphasize that the overall transition strength is not strictly temperature independent, but it decreases very slightly with increasing temperature. Since single-phonon excitations have a constant statistical factor,[29,30] this decrease arises solely from the change in the density of charged particles in the system, i.e., through the volume expansion factor $\mathcal{E}(T)$. The density of GaAs decreases by 0.25% when the temperature rises from 300 to 440 K [see Fig. 1(b)]. This expansion effect is, therefore, minor compared to the excitation energy shift and broadening, and the single-phonon transition strength can often be considered effectively temperature independent.

### D. Multi-phonon excitations

Multi-phonon excitations give rise to band absorption in the IR region, which is determined by the joint density of states (JDOS) of phonon excitations and by the probabilities of these excitation processes. At the critical points, the JDOS exhibits so-called van Hove singularities.[44] For crystalline silicon, two-phonon excitations can be modeled by a sophisticated approach[7,45] using Gaussian-broadened van Hove singularities describing additive and subtractive branches. The positions of the critical points are calculated using several phonon energies corresponding to the symmetries Γ, X, L, W, and K.

A similar model for GaAs will be developed in future work. However, in the present study, we use a somewhat simplified model that we have recently developed.[24]

The response function of multi-phonon excitations is modeled using 39 contributions,

$$\hat{\chi}_{\mathrm{mph}}(E,T)=\sum_{j=1}^{39}N_{\mathrm{mph},j}(T)\left[\hat{\beta}_{\mathrm{G}}*\mathcal{D}_{\mathrm{TDT},j}(E,T)\right],\tag{24}$$

where $\mathcal{D}_{\mathrm{TDT},j}(E,T)$ is the triangle–delta–triangle (TDT) dipole distribution function, $\hat{\beta}_{\mathrm{G}}$ is the Gaussian complex broadening function, and $*$ denotes the convolution operator. The dipole distribution function is nonzero in the vicinity of the excitation energy $E_j$, i.e., if the energies $E_j$ are sorted in the ascending order, then $E\in(E_{j-1},E_{j+1})$. The function is normalized with respect to the sum rule

$$\int_0^\infty E\left[\beta_{\mathrm{i,G}}*\mathcal{D}_{\mathrm{TDT},j}(E,T)\right]\mathrm{d}E=\int_{E_{j-1}}^{E_{j+1}}E\,\mathcal{D}_{\mathrm{TDT},j}(E,T)\,\mathrm{d}E=1,\tag{25}$$

where the invariance of the sum rule with respect to broadening was used.[46] A detailed description of the dipole distribution function together with Gaussian broadening is described in Appendix B. Moreover, Appendix C provides a simple example that facilitates understanding of the construction of the TDT dispersion model.

The transition strength of individual contributions, unlike that of single-phonon excitations, is strongly temperature dependent,

$$N_{\mathrm{mph},j}(T)=N_{\mathrm{mph},j}^{300\,\mathrm{K}}f_{\mathrm{mph}}^{0}\left(E_{\mathrm{mph},j}^{300\,\mathrm{K}},T\right),\tag{26}$$

where $f_{\mathrm{mph}}^{0}(E_{\mathrm{mph},j}^{300\,\mathrm{K}},T)$ is the temperature factor. This factor is related to the reference temperature $T_{\mathrm{R}}$, i.e., $f_{\mathrm{mph}}^{0}(E_j,T_{\mathrm{R}})=1$.

The temperature factor can be expressed as follows:

$$\begin{aligned}f_{\mathrm{mph}}^{0}(E_j,T)&=(1-t_j)f_{\mathrm{2ph}}^{0}(E_j,T)+t_jf_{\mathrm{3ph}}^{0}(E_j,T),\\ f_{\mathrm{2ph}}^{0}(E_j,T)&=\frac{f_{\mathrm{2ph}}(E_j,T)}{f_{\mathrm{2ph}}(E_j,T_{\mathrm{R}})},\\ f_{\mathrm{3ph}}^{0}(E_j,T)&=\frac{f_{\mathrm{3ph}}(E_j,T)}{f_{\mathrm{3ph}}(E_j,T_{\mathrm{R}})},\end{aligned}\tag{27}$$

i.e., as a superposition of the temperature factors of two- and three-phonon excitations. The parameter $t_j=0$ corresponds to pure two-phonon absorption, while $t_j=1$ corresponds to pure three-phonon absorption.

These temperature factors are approximated by additive statistical temperature factors of two- and three-phonon excitations with equally distributed phonon energies,

$$f_{\mathrm{2ph}}(E_j,T)=1+2f^{\mathrm{BE}}(E_j/2,T),\tag{28}$$

$$f_{\mathrm{3ph}}(E_j,T)=1+3f^{\mathrm{BE}}(E_j/3,T)+3\left[f^{\mathrm{BE}}(E_j/3,T)\right]^2,\tag{29}$$

where $f^{\mathrm{BE}}$ is the Bose–Einstein phonon statistical factor

$$f^{\mathrm{BE}}(E_{\mathrm{ph}},T)=\frac{1}{\exp\left(\frac{E_{\mathrm{ph}}}{k_{\mathrm{B}}T}\right)-1}.\tag{30}$$

This approximation is valid only within the temperature range of the experimental data (300–440 K), where the temperature factors can be assumed to vary linearly (two-phonon) or quadratically (three-phonon) with temperature,

$$f_{\mathrm{2ph}}^{0}(E_j,T)\approx\frac{T}{T_{\mathrm{R}}},\quad f_{\mathrm{3ph}}^{0}(E_j,T)\approx\left(\frac{T}{T_{\mathrm{R}}}\right)^2.\tag{31}$$

Similar to single-phonon absorption, the excitation energies $E_j$ are temperature dependent. It can be assumed that the individual phonon energies vary with temperature in the same way as TO phonons, therefore, we do not introduce an additional parameter $\varepsilon_{\mathrm{mph}}^{0\,\mathrm{K}}$ and instead use the same relation as in Eq. (17). The energies of multi-phonon excitations are, thus, parameterized by 40 parameters $E_{\mathrm{mph},j}^{300\,\mathrm{K}}$, where the parameter $E_{\mathrm{mph},40}^{300\,\mathrm{K}}$ defines the right boundary of the last triangular contribution. For the first contribution, we assume $E_0=0$.

The multi-phonon band broadening was parameterized by a single parameter $B_{\mathrm{mph},j}^{300\,\mathrm{K}}$ describing the Gaussian RMS value at 300 K. Its temperature dependence was described by the same relation (17) as for single-phonon excitations, with common parameters $\beta_{\mathrm{ph}}^{0\,\mathrm{K}}$ and Θ.

If the coupling between free carriers and single-phonon excitations is considered, it is reasonable to also assume coupling between multi-phonon excitations and free carriers. In an advanced model, it will be possible to consider coupling between multi-phonon and single-phonon excitations in a non-conducting sample, but this will require satisfying condition (14) for the entire coupled system. However, in the present work, this effect has been neglected, and only coupling for samples #2 and #3 was considered. This generalization was implemented by introducing a complex strength parameter in Eq. (24) through the following substitution:

$$N_{\mathrm{mph},j}^{300\,\mathrm{K}}\rightarrow N_{\mathrm{mph},j}^{300\,\mathrm{K}}+\mathrm{i}M_{\mathrm{mph},j}^{300\,\mathrm{K}}\frac{E_j}{E}.\tag{32}$$

In the TDT approach, each contribution is parameterized by four quantities describing the transition strength and the coupling:

$N^{300\,\mathrm{K}}_{\mathrm{mph,d},j}$ for the strength of the delta function, $N^{300\,\mathrm{K}}_{\mathrm{mph,l},j}$ for the left triangular part, $N^{300\,\mathrm{K}}_{\mathrm{mph,r},j}$ for the right triangular part, and $M^{300\,\mathrm{K}}_{\mathrm{mph},j}$ for the coupling, therefore,

$$N^{300\,\mathrm{K}}_{\mathrm{mph},j} = N^{300\,\mathrm{K}}_{\mathrm{mph,d},j} + N^{300\,\mathrm{K}}_{\mathrm{mph,l},j} + N^{300\,\mathrm{K}}_{\mathrm{mph,r},j}. \quad (33)$$

The coupling is divided among the individual terms in the same proportions as the strength, see Appendix B. These parameters, together with the remaining quantities describing the multi-phonon response, are summarized in Table VII.

The response function of sample #2 is constructed from the functions of samples #1 and #3 in the same manner as for the single-phonon absorption,

$$\hat{\chi}^{\#2}_{\mathrm{mph}}(E,T) = \left(1 - p^{\#2}_{\mathrm{mph}}\right)\hat{\chi}^{\#1}_{\mathrm{mph}}(E,T) + p^{\#2}_{\mathrm{mph}}\hat{\chi}^{\#3}_{\mathrm{mph}}(E,T). \quad (34)$$

This approach significantly reduces the number of free parameters. The value of $p^{\#2}_{\mathrm{mph}} = 0.0243(12)$ was found to be somewhat lower (approximately half) than the analogous parameter $p^{\#2}_{\mathrm{ph}} = 0.057(7)$.

**TABLE VII.** Dispersion parameters describing multi-phonon excitations.

| $j$ | $N^{300\,\mathrm{K}}_{\mathrm{mph,d},j}$ ($10^{-9}$ eV$^2$) | $N^{300\,\mathrm{K}}_{\mathrm{mph,l},j}$ ($10^{-9}$ eV$^2$) | $N^{300\,\mathrm{K}}_{\mathrm{mph,r},j}$ ($10^{-9}$ eV$^2$) | $E^{300\,\mathrm{K}}_{\mathrm{mph},j}$ (cm$^{-1}$) | $t_j$ | $M^{300\,\mathrm{K},\#3}_{\mathrm{mph},j}$ ($10^{-7}$ eV$^2$) |
|---|---|---|---|---|---|---|
| 1 | 69(7) | 523(13) | 508(36) | 73.9(2) | 0.23(3) | −596(114) |
| 2 | 0(19) | 391(18) | 889(40) | 97.1(14) | 0.04(2) | −1 366(123) |
| 3 | 38(4) | 781(35) | 859(28) | 129.3(5) | 0.11(2) | −1 027(144) |
| 4 | 1 517(1 395) | 2 203(131) | 0(2 170) | 161.2(10) | 0.05(2) | −188(61) |
| 5 | 469(1 639) | 1(3 386) | 2 330(7 320) | 167(9) | 0.00(7) | −70(117) |
| 6 | 9(11 345) | 1 465(1 382) | 797(20 487) | 182(33) | 0.00(9) | 66(149) |
| 7 | 4(2 579) | 268(11 206) | 2 952(10 400) | 186(30) | 0.00(9) | −44(221) |
| 8 | 1(14 942) | 2 616(7 929) | 800(67 243) | 206(56) | 0.00(11) | 163(1 638) |
| 9 | 1 703(65 544) | 2(91 345) | 5(105 169) | 211(41) | 0.0(4) | 73(3 700) |
| 10 | 2 035(42 074) | 3(115 796) | 0(17 653) | 219(35) | 0.0(3) | 85(2 613) |
| 11 | 2 283(17 737) | 1 236(12 701) | 0(40 804) | 227(17) | 0.00(15) | 83(565) |
| 12 | 5(8 480) | 1(20 893) | 9 940(5 960) | 233(8) | 0.0(2) | 261(433) |
| 13 | 11(3 480) | 6(1 826) | 1(366) | 258(2) | 1.0(6) | −1 248(2 275) |
| 14 | 139(571 563) | 0(2 160) | 4 553(566 212) | 267(111) | 0(3) | −14 941(114 566) |
| 15 | 5 442(115 540) | 3(111 133) | 6(324) | 269(5) | 1(2) | 10 768(110 865) |
| 16 | 400(939 021) | 13 363(323 409) | 2 044(1 379 009) | 287(215) | 0.00(10) | 2 248(29 449) |
| 17 | 8 270(1 205 654) | 2 601(1 349 571) | 1 567(68 811) | 289(35) | 0.00(6) | −1(33 266) |
| 18 | 5 757(90 905) | 2(200 986) | 9 620(20 591) | 294(16) | 0.00(5) | 885(5 199) |
| 19 | 3 618(2 247) | 7 129(4 456) | 16 447(1 252) | 308.9(10) | 0.00(4) | 1 505(268) |
| 20 | 710(507) | 5 450(336) | 6(2 230) | 341(2) | 0.00(6) | 1 077(237) |
| 21 | 1(6 913) | 1 591(26 411) | 1 009(19 539) | 349(70) | 0.00(5) | 669(262) |
| 22 | 819(1 569) | 1(3 401) | 0(2 721) | 357(9) | 0.00(10) | 300(987) |
| 23 | 412(3 114) | 0(5 835) | 1 566(465) | 361(5) | 0.03(3) | 641(850) |
| 24 | 0(8) | 805(153) | 745(24) | 387.9(3) | 0.10(3) | 1 228(127) |
| 25 | 240(44) | 1 080(43) | 1 018(174) | 409.4(4) | 0.07(2) | 724(52) |
| 26 | 0(65) | 1 035(128) | 978(220) | 425(2) | 0.22(4) | 591(45) |
| 27 | 26(173) | 1 331(394) | 2 054(427) | 439(2) | 0.00(4) | 612(34) |
| 28 | 341(68) | 1 056(205) | 1 589(280) | 454.6(6) | 0.22(4) | 577(54) |
| 29 | 0(154) | 1 058(61) | 502(430) | 475(3) | 0.32(4) | 483(49) |
| 30 | 136(99) | 338(576) | 1 443(1 961) | 484(4) | 0.08(5) | 367(381) |
| 31 | 0(1 207) | 1 575(2 330) | 216(4 205) | 504(27) | 0.2(3) | 215(454) |
| 32 | 321(1 372) | 1(3 560) | 1 812(1 809) | 507(9) | 0.02(9) | 296(95) |
| 33 | 2 264(518) | 1 194(418) | 521(958) | 523.5(5) | 0.00(7) | 198(37) |
| 34 | 82(325) | 900(1 844) | 677(2 170) | 535(9) | 0.08(6) | 212(46) |
| 35 | 112(321) | 0(716) | 583(196) | 541(4) | 0.28(5) | 181(81) |
| 36 | 0(5) | 111(27) | 122(31) | 560.9(10) | 0.87(9) | 227(40) |
| 37 | 18(55) | 101(27) | 0(131) | 581(5) | 0.57(11) | 72(41) |
| 38 | 0(107) | 16(1 036) | 114(663) | 586(155) | 1.0(7) | 165(129) |
| 39 | 0(10) | 60(238) | 34(23) | 625(11) | 0.6(2) | 52(141) |

$B^{300\,\mathrm{K}}_{\mathrm{mph},j} = 3.63(11)\,\mathrm{cm}^{-1}$, $E^{300\,\mathrm{K}}_{\mathrm{mph},40} = 654.6(13)\,\mathrm{cm}^{-1}$, $\varepsilon^{0\,\mathrm{K}}_{\mathrm{ph}} = 1.012\,03(7)$, $\beta^{0\,\mathrm{K}}_{\mathrm{ph}} = 0.890(8)$, $\Theta = 286(2)$ K, $p^{\#2}_{\mathrm{mph}} = 0.0243(12)$

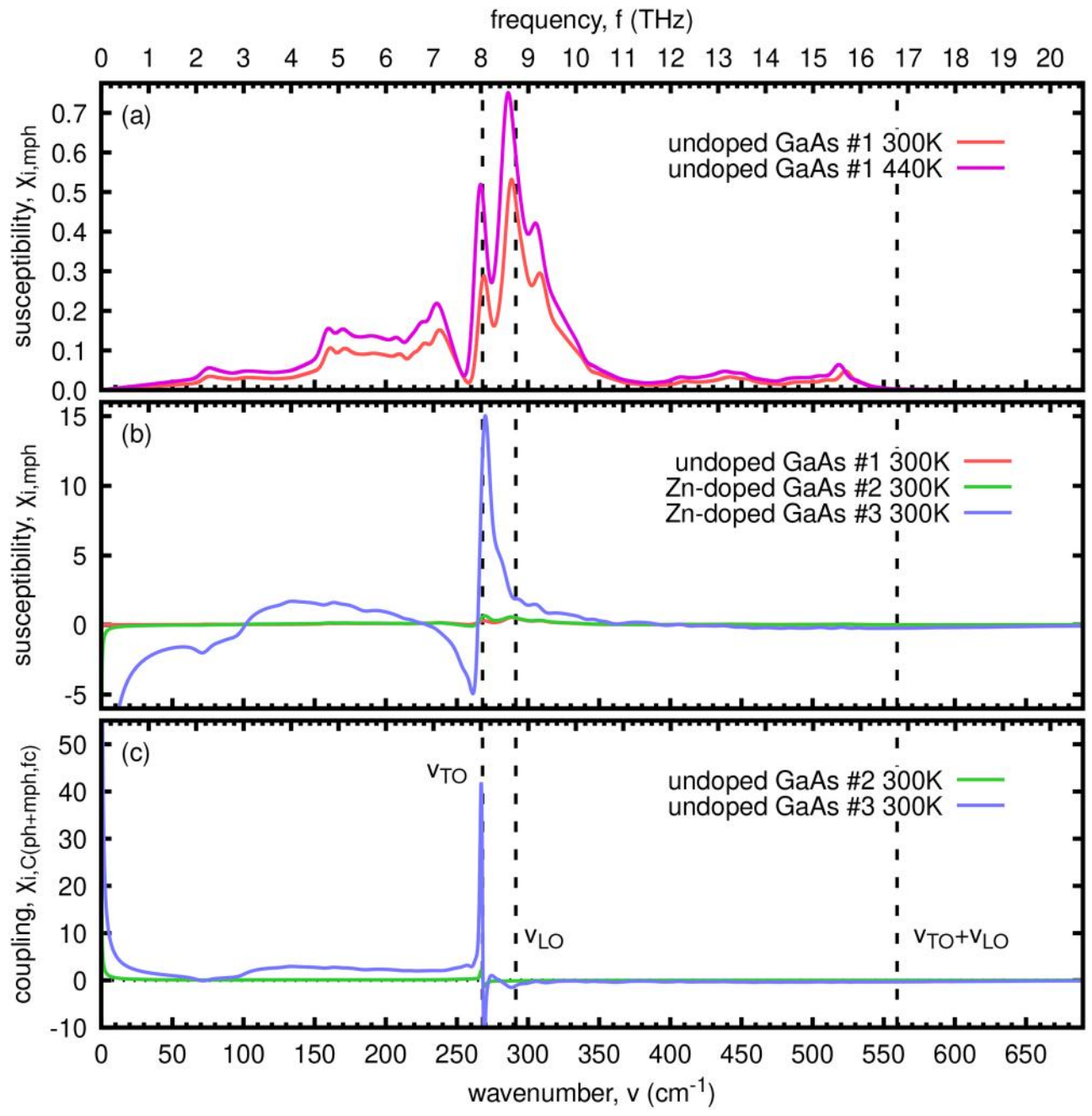

**FIG. 11.** Spectral dependence of the imaginary part of the susceptibility corresponding to multi-phonon excitations: (a) dependence on temperature; (b) dependence on Zn concentration. (c) Interaction term expressing the coupling of all phonon excitations with free carriers. The TO and LO frequencies were determined in this work by Eqs. (19) and (36).

Figure 11(a) illustrates the resulting temperature dependencies of the imaginary part of the response function. Beyond the reststrahlen region, the absorption spectrum exhibits a classical multi-phonon pattern, similar to that observed in crystalline silicon. Most structures exhibit a temperature dependence corresponding to two-phonon absorption (i.e., an approximately linear increase of their transition strength with temperature), except for the region between 400 and 475 $\text{cm}^{-1}$. This linear temperature dependence reflects the leading-order contribution of two-phonon processes governed by Bose–Einstein occupation factors. In contrast, the spectral region between 400 and 475 $\text{cm}^{-1}$ deviates from this behavior. This is evident from the values of the parameter $t_j$ listed in Table VII, where positive values of $t_j$ indicate a nonlinear, predominantly quadratic temperature dependence of the absorption strength. Such a nonlinear dependence is characteristic of a mixed contribution of two- and three-phonon excitations, see the definition of the parameter $t_j$ in (27). In the reststrahlen region, the determination of optical constants is subject to a large error, but if the results are to be believed, it is possible that this region is strongly influenced by the phonon–phonon interaction.

Figure 11(b) shows the dependence on the dopant concentration. Here, the Fano resonance effect is clearly visible in the region around the TO frequency. This indicates a strong coupling between the two-phonon part of the system and the TO phonon-induced presence of free carriers. The system must be viewed as a whole, therefore, in Fig. 11(c), the interaction part corresponding to the interaction of free carriers with the complete phonon spectrum was shown.

The excitation part of the response function of multi-phonon excitations compared to the interaction part is almost negligible, so the response function of the heavily doped sample practically represents the interaction part. It should be emphasized that Figs. 10–12 are based on the same final dispersion model; only different components or combinations of the phonon response functions are shown for clarity. Comparing Figs. 10(b) and 11(b), it is clear that the Fano resonance effects of single-phonon and multi-phonon excitations at the TO frequency oppose each other and partially cancel, resulting in the interference spectrum in Fig. 11(c), representing phonon–plasmon coupling. While Fig. 10 shows only the single-phonon response of the three studied samples, Fig. 12 presents the total phonon response, i.e., the sum of single-phonon and multi-phonon contributions. A detail of the coupling functions in Fig. 11(c) is shown in Fig. 12(b). This representation is more robust, as the separation of single-phonon and multi-phonon contributions relies solely on their different temperature dependences and is, therefore, intrinsically ambiguous, especially when performed over a relatively limited temperature range. Extending the temperature range in future studies could improve the separability of single-phonon and multi-phonon excitations and allow a more reliable decomposition of the total phonon response.

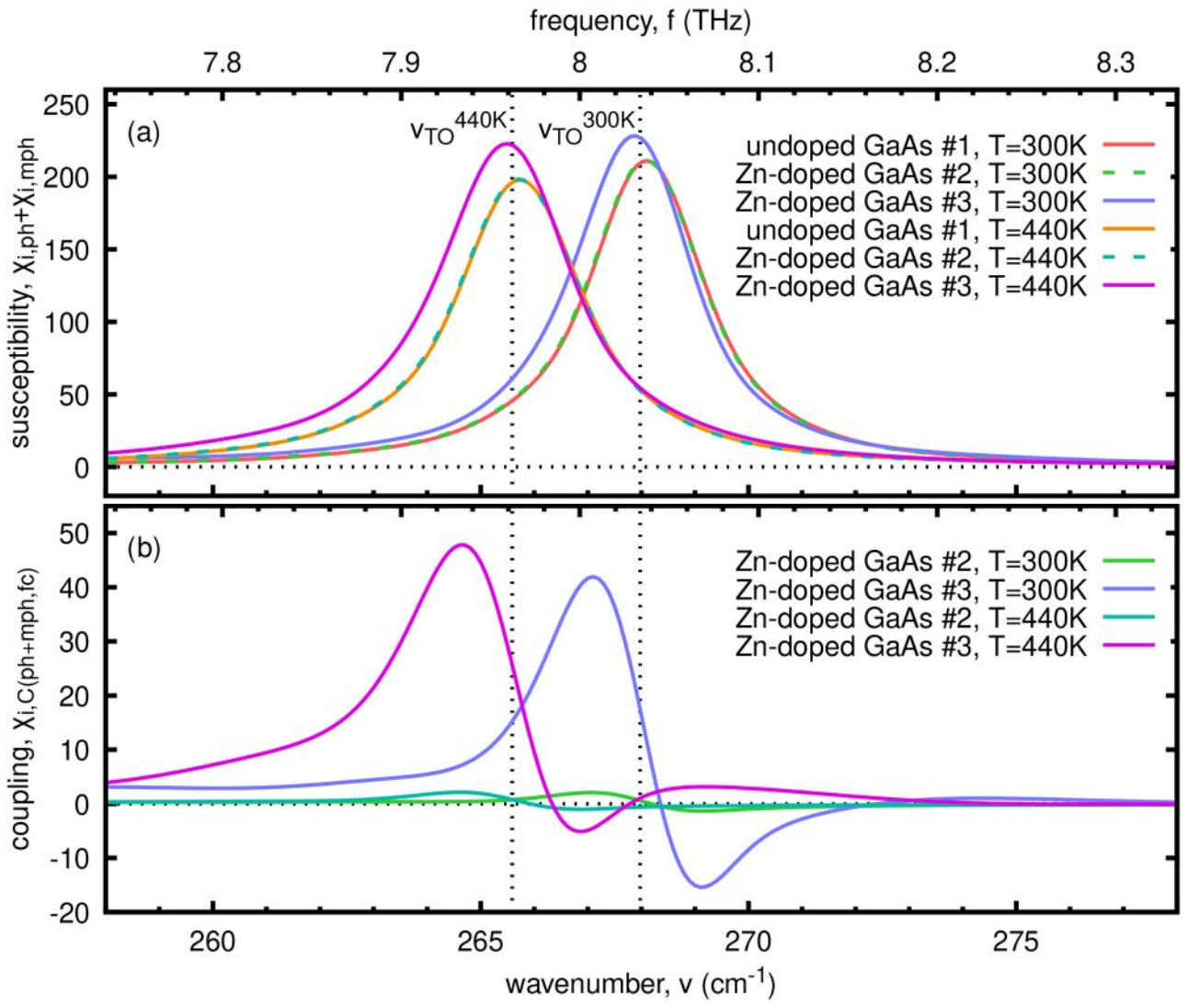

**FIG. 12.** (a) Spectral dependence of the imaginary part of the susceptibility of total phonon excitations as a function of temperature and Zn concentration. (b) Interaction term expressing the coupling of all phonon excitations with free carriers [curves for 300 K are a detail of Fig. 11(c)]. Vertical dotted lines indicate the mean TO-phonon resonance frequency at 300 and 440 K, calculated as a weighted mean using (19).

Despite the relatively large uncertainties in the individual parameters, the resulting optical constants (see Fig. 4) remain stable. This behavior indicates strong over-parameterization of the model and substantial correlations between parameters. The largest uncertainties occur in the spectral interval where the samples are opaque, i.e., in the reststrahlen region.

At the present stage of model development, we are unable to further reduce these correlations. It is possible that the situation will improve once an advanced dispersion model—incorporating a limited set of critical points to better reflect van Hove singularities—is implemented. However, it is also conceivable that strong correlations are an intrinsic feature of coupled systems, making the problem more fundamental.

Figure 13 compares the phonon absorption in GaAs obtained using the classical Lorentzian model, its Voigt-profile modification, and the final model introduced in this work, which describes the phonon absorption as a combination of single-phonon and multi-phonon contributions. The comparison demonstrates why models containing Lorentzian components do not allow accurate modeling of sample transmittance in spectral regions where the samples exhibit partial transparency. In the spectral ranges 220–250, 350–400, and above 550 $\mathrm{cm}^{-1}$, the actual absorption is lower than predicted by these models. Thus, the classical model provides a valid description only in the immediate vicinity of the single-phonon resonance frequency.

As seen in Fig. 13, the single-phonon contribution of our model attains negative values on both sides of the resonant frequency. This behavior is a consequence of the fact that the response is composed of several mutually coupled phonon excitations. Such coupling can naturally arise because gallium occurs in isotopes with different atomic masses, which may exhibit slightly different resonance frequencies and respond with different phases to an external harmonic field. In this case, the resulting phase shift

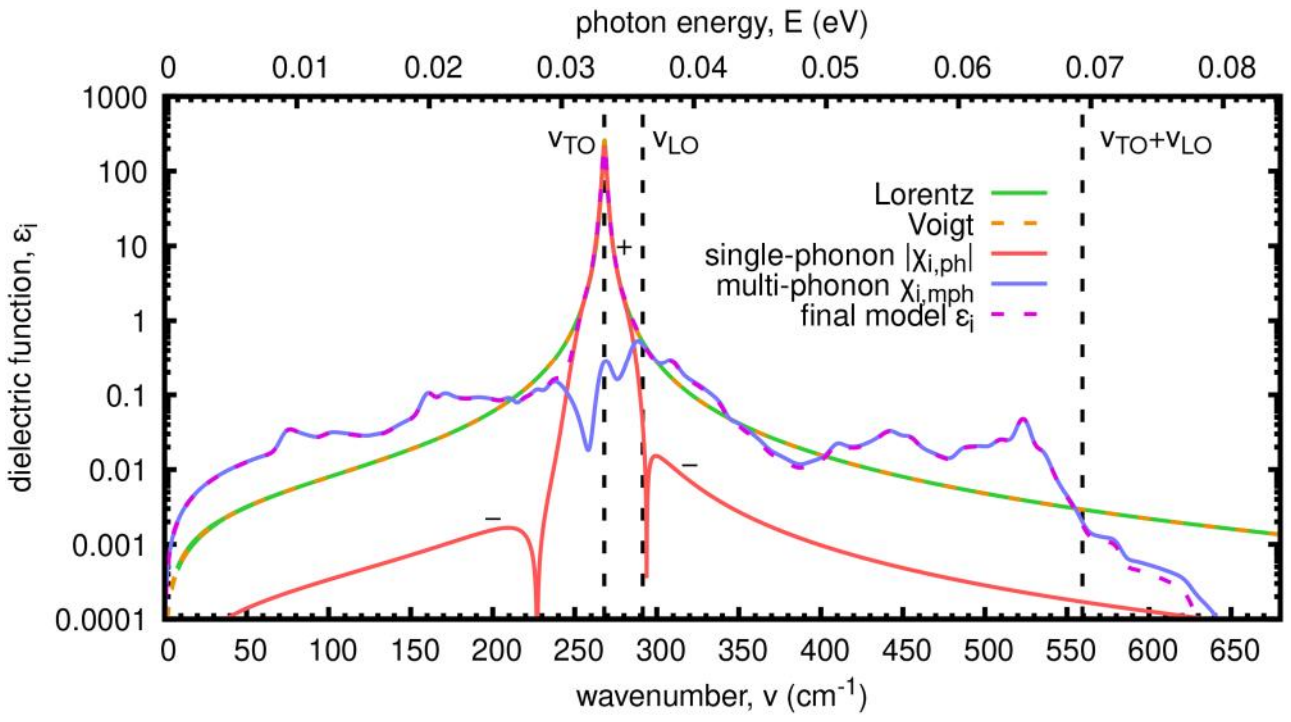

**FIG. 13.** Comparison of the imaginary part of the dielectric function calculated using the classical model (Lorentz), the model employing the Voigt distribution, and the final dispersion model (1), which treats the phonon contribution using only Gaussian-type broadening. Because the single-phonon response includes coupling, the function takes both positive and negative values, which are explicitly indicated. The TO and LO frequencies were determined in this work using Eqs. (19) and (36).

is opposite to that expected for a system composed of gallium atoms with identical masses.[24]

The situation could change if the coupling between single-phonon and multi-phonon excitations were introduced for sample #1. In that case, resonance effects could render the multi-phonon response function negative on the background of the single-phonon excitations, thereby allowing for a more slowly decaying response function of the single-phonon contribution. However, the dispersion model is already highly correlated, and introducing additional independent parameters would further exacerbate this issue.

In the future, the situation could be improved by extending the temperature range of the experimental data because the temperature dependence of the excitation strength helps separate the phonon absorption into single- and multi-phonon contributions.

The total transition strength of multi-phonon excitations is

$$N_{\mathrm{mph}}^{300\,\mathrm{K}} = \sum_{j=1}^{39} \sum_{q=\mathrm{c,l,r}} N_{\mathrm{mph},q,j}^{300\,\mathrm{K}} = 0.1559 \times 10^{-3}\,\mathrm{eV}^2. \tag{35}$$

If this strength is added to the strength of single-phonon excitations (20) and used in Eq. (23), the following value of the LO-phonon energy is obtained:

$$E_{\mathrm{LO}} = \sqrt{E_{\mathrm{TO}}^2 + \frac{2}{\pi}\frac{N_{\mathrm{ph}} + N_{\mathrm{mph}}}{\varepsilon_\infty}} = 36.1297\,\mathrm{meV} \tag{36}$$

(291.4 $\mathrm{cm}^{-1}$, 8.7361 THz).

This value corresponds precisely to the maximum of the loss function, i.e., the LO-phonon frequency. This can be expected because in the classical model, the whole transition strength is concentrated in the single-phonon excitations and multi-phonon excitations do not appear at all, while the real system is more complex. In the final model, which is based on quantum-mechanical approaches, the strength is distributed between single-phonon and multi-phonon contributions. Therefore, when applying the relation (23) derived from the classical HO model, it is necessary to use the total polarizability of the lattice, i.e., the total transition strength of all phonon excitations.

Using the TO- and LO-phonon energies, the theoretical maximum energy of two-phonon excitations can be defined because the LO+TO branch has its maximum energy at the Γ point,

$$E_{2\mathrm{ph}}^{\max} = E_{\mathrm{TO}} + E_{\mathrm{LO}} = 69.3548\,\mathrm{meV} \tag{37}$$

(559.38 $\mathrm{cm}^{-1}$, 16.77 THz).

Above this energy, only three-phonon and higher-order excitations with much lower probability can occur. This cutoff is clearly visible in the transmission spectrum of undoped GaAs #1 (see Figs. 13 and 14).

From the comparison of the parameters $M_{\mathrm{mph},j}^{300\,\mathrm{K},\#3}$ and $N_{\mathrm{mph},j}^{300\,\mathrm{K}}$, it is clear that the asymmetric part describing the coupling with free carriers is about two orders of magnitude stronger than the symmetric part. This is also evident in Fig. 11(b), where the imaginary part of susceptibility of sample #1 is negligible compared to that of sample #3. At the same time, the sum-rule integral of sample #3 is essentially the same as those of samples #1 and #2, i.e., in Eq. (35), it is almost zero relative to the values of $\chi^{\#3}_{\mathrm{i,mph}}(E, T)$. This occurs because the transition strength function $E\chi_{\mathrm{i,mph}}(E)$ takes both positive and negative values in the frequency domain, and the negative contributions nearly cancel the positive ones. Note that, unlike the excitation strengths $N^{300\,\mathrm{K},\#3}_{\mathrm{mph},j}$, the f-sum-rule integral does not depend on the values of $M^{300\,\mathrm{K},\#3}_{\mathrm{mph},j}$ since the corresponding interaction contribution is antisymmetric in the frequency domain and, therefore, integrates exactly to zero.

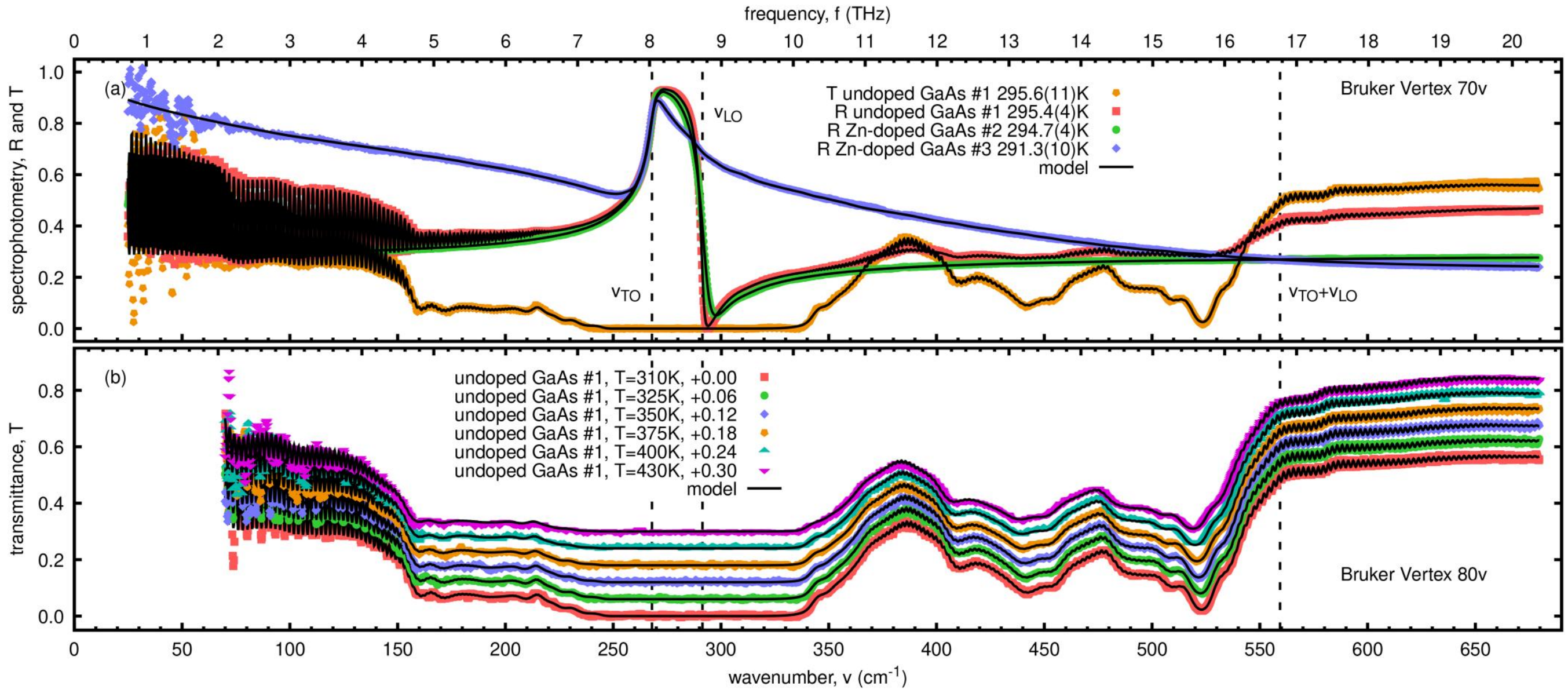

**FIG. 14.** Selected experimental data in the region of multi-phonon excitations in GaAs. The vertical dashed lines indicate, from left to right, the resonant TO- and LO-phonon frequencies and their sum, which determines the upper limit of two-phonon excitations. The frequencies were determined in this work using Eqs. (19) and (36).

In the time domain, this effect corresponds to a weak subsystem oscillating out of phase with the mean field, with a phase shift close to $\pm 90°$. This part contributes only minimally to the total polarizability of the medium because another part of the system compensates this out-of-phase motion. This phenomenon is known as a Fano resonance,[47] in which weak excitations respond to a strong background via the local field rather than the mean field. In the frequency domain, instead of a peak, S-shaped functions—i.e., the Hilbert transforms of peaks—are observed.

## E. Free carriers

Zn-doped GaAs crystals are p-type semiconductors. The top of the valence band consists of two subbands whose maxima are degenerate at the Γ point. These subbands have different effective masses, so contributions from light and heavy holes can be distinguished in the optical spectra.

In this work, free carriers are modeled using the classical dispersion approach, which is sufficient for the spectral range of the experimental data considered. It is further assumed that the hole density remains constant across the temperature range studied, with only the carrier collision probability, i.e., the broadening parameter, changing with temperature.

As shown in our theoretical study,[24] coupling between free carriers cannot be considered. Consequently, light and heavy holes are described using two separate Drude terms,

$$\hat{\chi}_{\mathrm{fc}}(E, T) = -\frac{2}{\pi}\sum_{j=1}^{2}\frac{N_{\mathrm{fc},j}}{E^2 + \mathrm{i}E\Gamma_{\mathrm{fc},j}(T)}. \tag{38}$$

Table VIII summarizes the parameters describing free carriers in doped GaAs. The ratio of the transition strengths $N^{\#2}_{\mathrm{fc},j}/N^{\#3}_{\mathrm{fc},j}$, corresponding to light and heavy holes, allows estimation of the doping ratio between samples #2 and #3. For light holes ($j = 1$), this ratio is 5.47%, in excellent agreement with the value $p^{\#2}_{\mathrm{ph}} = 5.67\%$ obtained from the single-phonon excitation model. For heavy holes ($j = 2$), which contribute less prominently at lower energies, the ratio is 6.17%, again showing good agreement.

**TABLE VIII.** Dispersion parameters describing free carriers.

| $j$ | $N^{\#2}_{\mathrm{fc},j}$ ($10^{-3}$ eV$^2$) | $\Gamma^{300\,\mathrm{K},\#2}_{\mathrm{fc},j}$ (eV) | $N^{\#3}_{\mathrm{fc},j}$ ($10^{-3}$ eV$^2$) | $\Gamma^{300\,\mathrm{K},\#3}_{\mathrm{fc},j}$ (eV) |
|---|---|---|---|---|
| 1 | 5.46(11) | 0.134(4) | 99.8(4) | 0.1451(10) |
| 2 | 1.73(2) | 0.012 63(10) | 28.0(5) | 0.0271(5) |

$\beta^{0\,\mathrm{K},\#2}_{\mathrm{fc}} = 0.546(9)$, $\beta^{0\,\mathrm{K},\#3}_{\mathrm{fc}} = 0.690(7)$, $\Theta = 286(2)$ K

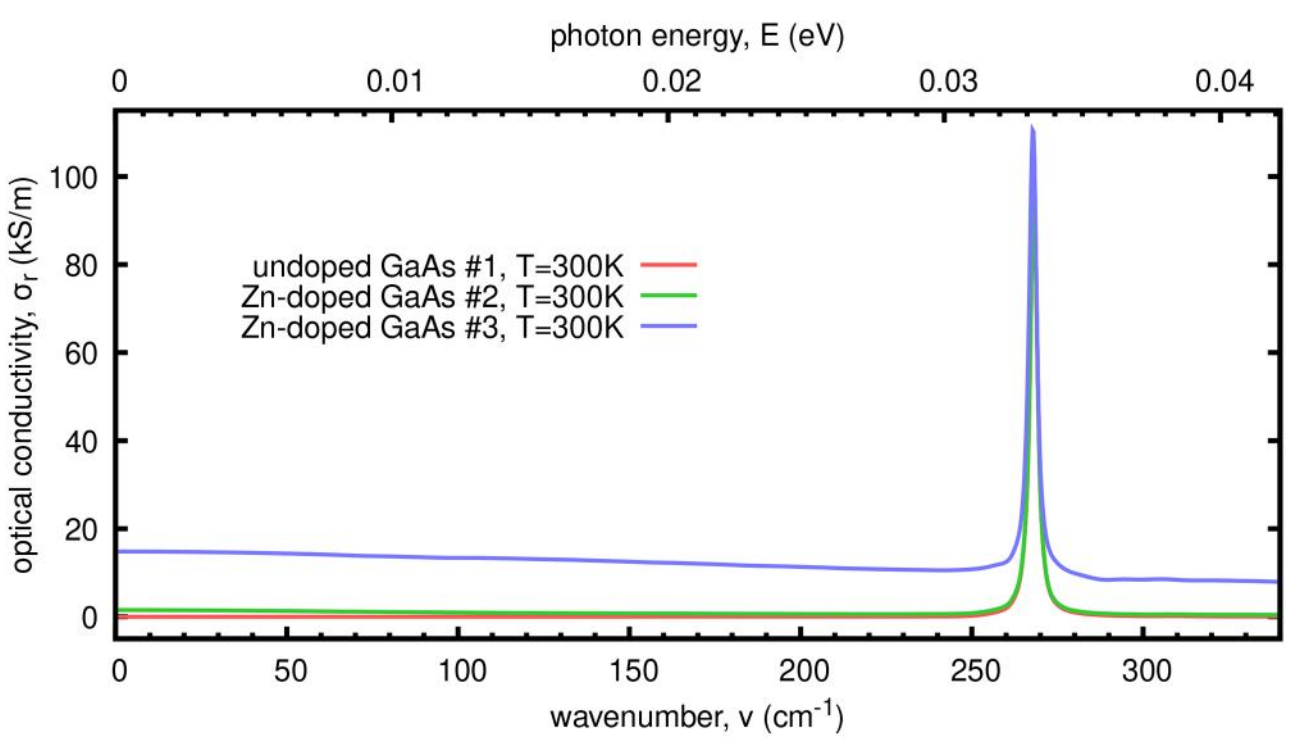

**FIG. 15.** Real part of the optical conductivity of GaAs samples.

Figure 15 presents the real part of the optical conductivity, with the response function defined as

$$\hat{\sigma}(E) = -\mathrm{i}\frac{\epsilon_0}{\hbar}\, E\,[\hat{\varepsilon}(E) - 1]. \tag{39}$$

This representation is convenient because it has a finite static value corresponding to the material's electrical conductivity, $\sigma_0 = \sigma_r(0)$. The static conductivity values obtained from optical characterization are listed in Table I and compared with the nominal values reported by the wafer manufacturer.

Within the range of temperatures that was the subject of the study, it was assumed that the temperature dependence of free carriers is given by the temperature dependence of their broadening parameters. Consequently, the temperature-dependent parameters were modeled in accordance with the Bose–Einstein-type dependence, as implemented in other parts of the system.

However, in a wider temperature range, the previously defined temperature dependence with a constant free-carrier concentration is not sufficient. The construction of an accurate model based on Fermi–Dirac statistics would be needed.

## V. CONCLUSION

In this work, it was applied for the first time a recently published approach[24,40] to describe the coupling between free carriers and phonon excitations in the zinc blende cubic crystal. Modeling phonon excitations required replacing the classical Lorentzian broadening with multiple Gaussian broadenings. To capture specific absorption features in the transparent region of the GaAs wafer, multi-phonon absorption was included using Gaussian-broadened bands constructed *ad hoc* from triangle–delta–triangle distribution functions. The phonon part of the dispersion model was constructed by separating single-, two-, and three-phonon contributions based on their different temperature dependences, primarily in the vicinity of the TO-phonon resonance where these contributions spectrally overlap. Although the investigated temperature range was relatively limited, the use of temperature-dependent data was essential for this separation, as a model based solely on room-temperature measurements would not allow such a decomposition. At the same time, we acknowledge that a wider temperature range would allow this separation to be tested and refined further.

Free carriers in the far-IR region were described classically with two Drude terms representing light and heavy holes in Zn-doped GaAs wafers. It has been shown that increasing dopant density leads to a significant asymmetry of the TO-phonon peak, representing a classical manifestation of the coupling between bound and free excitations and its impact on the low-frequency response of the system. The asymmetry of the peak is consistent with the Fano resonance effect,[47] where sharp excitations on a broad absorption background induce quantum interference. This resonance effect has also been applied to absorption bands describing multi-phonon excitations. Within this modeling framework, the static conductivity is not introduced as an independent quantity but emerges directly from the coupling between bound and free excitations.

The main purpose of this pilot work was to test new approaches in modeling coupled systems in the infrared region using models providing physically consistent response functions over a wide spectral and temperature range.

Although the dispersion model is still under development, the resulting temperature-dependent optical+ constants of GaAs represent a level of accuracy not previously available.

## ACKNOWLEDGMENTS

This research was supported by the projects CEPLANT (No. LM2023039), CzechNanoLab (No. LM2023051) Research Infrastructures, and LasApp (No. CZ.02.01.01/00/22_008/0004573) funded by the Ministry of Education, Youth and Sports of the Czech Republic. The work at NMSU was supported by the Air Force Office of Scientific Research under Award No. FA9550-24-1-0061.

## AUTHOR DECLARATIONS

### Conflict of Interest

The authors have no conflicts to disclose.

### Author Contributions

**Daniel Franta:** Conceptualization (lead); Formal analysis (lead); Methodology (equal); Software (equal); Visualization (lead); Writing – original draft (lead). **Beáta Hroncová:** Methodology (equal); Software (equal); Writing – original draft (supporting); Writing – review & editing (equal). **Mihai-George Mureşan:** Conceptualization (supporting); Writing – original draft (supporting); Writing – review & editing (equal). **Stefan Zollner:** Conceptualization (supporting); Funding acquisition (equal); Supervision (equal); Writing – original draft (supporting); Writing – review & editing (supporting).

## DATA AVAILABILITY

The data that support the findings of this study are available from the corresponding author upon reasonable request and will also be made openly available in the RefractiveIndex.info database, Ref. 48.

## APPENDIX A: SINGLE-PHONON DISPERSION MODELS OF UNCOUPLED TO PHONONS

If a non-conducting crystal exhibits a single non-interacting TO phonon, the dielectric function in the vicinity of this phonon can be accurately described by the classical Lorentz model, traditionally written as

$$\hat{\varepsilon}(E) = 1 + \hat{\chi}_{\rm el}(E) + \frac{E_{\rm p}^2}{E_{\rm c}^2 - E^2 - {\rm i}\Gamma E}, \tag{A1}$$

where $\hat{\chi}_{\rm el}(E)$ is the electronic response function, $E_{\rm c}$ is the central frequency, $E_{\rm p}$ is the plasma frequency, and $\Gamma$ denotes the damping (broadening) parameter, corresponding to the full width at half maximum of the absorption peak.

This dielectric function corresponds to the dispersion model (1) evaluated at the reference temperature, provided that multi-phonon excitations are neglected and the volume expansion factor is set to unity. For narrow absorption peaks, the central frequency approximately corresponds to the TO-phonon resonance energy $E_{\rm TO}$ according to

$$E_{\rm c}^2 = E_{\rm TO}^2 + \Gamma^2/4. \tag{A2}$$

Instead of the square of the plasma frequency, the amplitude of the response function is often used. However, our models are parameterized using the f-sum rule

$$\int_0^\infty E\,\varepsilon_{\rm i}(E)\,{\rm d}E = \frac{\pi}{2}E_{\rm p}^2 = N, \tag{A3}$$

where $N$ is the transition strength parameter, which is proportional to the oscillator amplitude, i.e., to the density of charged particles, up to the factor $\pi/2$.

Our parameterization of the Lorentz model then has the form[40]

$$\hat{\varepsilon}(E) = \varepsilon_\infty + \frac{2N/\pi}{E_{\rm TO}^2 + \Gamma^2/4 - E^2 - {\rm i}\Gamma E}, \tag{A4}$$

where $1 + \hat{\chi}_{\rm el}(E)$ is approximated by the constant $\varepsilon_\infty$, which is justified because the electronic excitation energies are much higher than the TO-phonon energies.

To distinguish this parameterization from the classical Lorentz model, we refer to it as a Lorentzian-broadened discrete excitation.

An advantage of the classical Lorentz model is that it can be rewritten in a factorized form

$$\hat{\varepsilon}(E) = \varepsilon_\infty \frac{E_{\rm LO}^2 + \Gamma^2/4 - E^2 - {\rm i}\Gamma E}{E_{\rm TO}^2 + \Gamma^2/4 - E^2 - {\rm i}\Gamma E}, \tag{A5}$$

where $E_{\rm TO}$ and $E_{\rm LO}$ are the transverse and longitudinal optical phonon resonance energies, respectively, defined consistently with the Lorentzian-broadened parameterization used above. In particular, $E_{\rm LO}$ corresponds to the position of the absorption maximum in the loss function $L_{\rm r}(E)$, where $\hat{L}(E) = -{\rm i}\hat{\varepsilon}^{-1}(E)$.

The classical model can be generalized by replacing the Lorentzian line shape with a Gaussian function or a Voigt profile, which is the convolution of Lorentzian and Gaussian functions.

The Gaussian-broadened phonon excitation is modeled by the following expressions:[40]

$$\begin{aligned}\hat{\varepsilon}(E) = \varepsilon_\infty &+ \frac{-\sqrt{2}N}{\pi B E_{\rm TO}}\left[{\rm D}\left(\frac{E - E_{\rm TO}}{\sqrt{2}B}\right) - {\rm D}\left(\frac{E + E_{\rm TO}}{\sqrt{2}B}\right)\right] \\ &+ \frac{{\rm i}N}{\sqrt{2\pi}BE_{\rm TO}}\left[\exp\left(-\frac{(E - E_{\rm TO})^2}{2B^2}\right) - \exp\left(-\frac{(E + E_{\rm TO})^2}{2B^2}\right)\right],\end{aligned} \tag{A6}$$

where the broadening parameter $B$ represents the RMS width of the Gaussian peak.

The Dawson function[41,42] is defined as

$${\rm D}(x) = \exp(-x^2)\int_0^x \exp(t^2)\,{\rm d}t. \tag{A7}$$

Up to a constant factor, the Dawson function is the Hilbert transform of the Gaussian function, thus ensuring the Kramers–Kronig consistency of the model.

A generalization to a Voigt absorption profile can be written compactly using the complex Faddeeva function[40]

$$\hat{\varepsilon}(E) = \varepsilon_\infty + \frac{{\rm i}N}{\sqrt{2\pi}BE_{\rm TO}}\left[\hat{\rm W}\left(\frac{E - E_{\rm TO} + {\rm i}\Gamma/2}{\sqrt{2}B}\right) - \hat{\rm W}\left(\frac{E + E_{\rm TO} + {\rm i}\Gamma/2}{\sqrt{2}B}\right)\right]. \tag{A8}$$

The Faddeeva function of a complex argument is defined by

$$\hat{\rm W}(\hat{z}) = \frac{\rm i}{\pi}\int_{-\infty}^{\infty}\frac{\exp(-t^2)}{\hat{z} - t}\,{\rm d}t. \tag{A9}$$

From a computational point of view, Eq. (A8) reduces to Eq. (A4) in the limit $B \to 0$.

Note that in this work, we prefer the parameterization using $E_{\rm TO}$ rather than $E_{\rm c}$ as it has a clearer physical interpretation. From a numerical point of view, however, the difference between $E_{\rm c}$ and $E_{\rm TO}$ is usually negligible. For the GaAs parameters listed in Table V, the difference is within the statistical uncertainty

$$E_{\rm c} - E_{\rm TO} = 0.000\,256\,7\,{\rm meV}.$$

## APPENDIX B: TRIANGLE–DELTA–TRIANGLE (TDT) DISPERSION MODEL

The distribution function of the $j$ th element of the TDT dispersion model is defined on the interval from $E_{j-1}$ to $E_{j+1}$ as follows:

$$\begin{aligned}\mathcal{D}_{\mathrm{TDT},j}(E,T) &= \mathrm{sgn}(E)[A_{\mathrm{l},j}\,\Pi_{E_{j-1},E_j}(|E|)\,(|E|-E_{j-1}(T)) \\ &\quad + A_{\mathrm{d},j}\delta(|E|-E_j(T)) \\ &\quad + A_{\mathrm{r},j}\,\Pi_{E_j,E_{j+1}}(|E|)\,(E_{j+1}(T)-|E|)].\end{aligned} \tag{B1}$$

The cutoff function $\Pi_{a,b}(\cdot)$ is equal to unity on the interval from $a$ to $b$ and vanishes outside this interval. It can be defined using two Heaviside functions $\Theta(\cdot)$ as

$$\Pi_{a,b}(x) = \Theta(x-a)\Theta(b-x). \tag{B2}$$

The distribution function is normalized with respect to the sum rule (25), and the ratio of the individual triangle–delta–triangle parts is given by (33), which defines the amplitudes of the respective contributions as

$$A_{\mathrm{l},j} = \frac{6N^{300\,\mathrm{K}}_{\mathrm{mph,l},j}}{N^{300\,\mathrm{K}}_{\mathrm{mph},j}\left(2E_j^3 - 3E_j^2E_{j-1} + E_{j-1}^3\right)}, \tag{B3}$$

$$A_{\mathrm{d},j} = \frac{N^{300\,\mathrm{K}}_{\mathrm{mph,d},j}}{N^{300\,\mathrm{K}}_{\mathrm{mph},j}E_j}, \tag{B4}$$

$$A_{\mathrm{r},j} = \frac{6N^{300\,\mathrm{K}}_{\mathrm{mph,r},j}}{N^{300\,\mathrm{K}}_{\mathrm{mph},j}\left(2E_j^3 - 3E_j^2E_{j+1} + E_{j+1}^3\right)}. \tag{B5}$$

For brevity, the explicit temperature dependence of the energies, $E_j \equiv E_j(T)$, has been omitted in the above expressions.

To express the normalized contribution to the response function, each term must be broadened, i.e., convolved with a complex broadening function

$$\begin{aligned}\hat{\beta}_{\mathrm{G}} * \mathcal{D}_{\mathrm{TDT},j}(E,T) &= \int_{-\infty}^{\infty} \hat{\beta}_{\mathrm{G}}(E-X)\,\mathcal{D}_{\mathrm{TDT},j}(X,T)\,\mathrm{d}X \\ &= \hat{\chi}^0_{\mathrm{l},j}(E,T) + \hat{\chi}^0_{\mathrm{d},j}(E,T) + \hat{\chi}^0_{\mathrm{r},j}(E,T),\end{aligned} \tag{B6}$$

where the complex Gaussian broadening function containing the parameter $B$ is given by[40]

$$\hat{\beta}_{\mathrm{G}}(x) = -\frac{\sqrt{2}}{\pi B}\mathrm{D}\left(\frac{x}{\sqrt{2}B}\right) + \frac{\mathrm{i}}{\sqrt{2\pi}B}\exp\left(-\frac{x^2}{2B^2}\right). \tag{B7}$$

In fact, the broadening parameter is also temperature dependent, $B \equiv B(T)$, so the normalized distribution function depends on temperature through both the $E_j$ and $B$ parameters.

While the convolution with a delta function is straightforward, i.e.,

$$\begin{aligned}\hat{\chi}^0_{\mathrm{d},j}(E,T) &= \frac{-\sqrt{2}A_{\mathrm{d},j}}{\pi B}\left[\mathrm{D}\left(X_j^-\right) - \mathrm{D}\left(X_j^+\right)\right] \\ &\quad + \frac{\mathrm{i}A_{\mathrm{d},j}}{\sqrt{2\pi}B}\left[\exp\left(-\left(X_j^-\right)^2\right) - \exp\left(-\left(X_j^+\right)^2\right)\right],\end{aligned} \tag{B8}$$

where

$$X_j^{\pm} = \frac{E \pm E_j}{\sqrt{2}B}, \tag{B9}$$

the convolution with the linear functions describing the left and right triangular parts requires the introduction of an additional special function, namely, the integral of the Dawson function[40,49]

$$\mathrm{D_i}(x) = \int_0^x \mathrm{D}(t)\,\mathrm{d}t. \tag{B10}$$

For completeness, we recall the definition of the error function

$$\mathrm{erf}(x) = \frac{2}{\sqrt{\pi}}\int_0^x \exp(-t^2)\,\mathrm{d}t, \tag{B11}$$

which is also necessary for calculations.

The normalized susceptibilities corresponding to the left and right triangles are calculated as follows:

$$\begin{aligned}\hat{\chi}^0_{\mathrm{l},j}&(E,T) \\ &= A_{\mathrm{l},j}\frac{2}{\pi}\left[(E_j - E_{j-1}) + \frac{B}{\sqrt{2}}L_{\mathrm{D}}(E) + L_{\mathrm{D_i}}(E)\right] \\ &\quad + \mathrm{i}A_{\mathrm{l},j}\left[\frac{1}{2}L_{\mathrm{erf}}(E) + \frac{B}{\sqrt{2\pi}}L_{\mathrm{G}}(E)\right]\end{aligned} \tag{B12}$$

and

$$\begin{aligned}\hat{\chi}^0_{\mathrm{r},j}&(E,T) \\ &= A_{\mathrm{r},j}\frac{2}{\pi}\left[(E_j - E_{j+1}) + \frac{B}{\sqrt{2}}R_{\mathrm{D}}(E) + R_{\mathrm{D_i}}(E)\right] \\ &\quad + \mathrm{i}A_{\mathrm{r},j}\left[\frac{1}{2}R_{\mathrm{erf}}(E) + \frac{B}{\sqrt{2\pi}}R_{\mathrm{G}}(E)\right],\end{aligned} \tag{B13}$$

where

$$L_{\rm G}(E) = \exp\left[-\left(X_{j-1}^{-}\right)^2\right] - \exp\left[-\left(X_{j}^{-}\right)^2\right] + \exp\left[-\left(X_{j}^{+}\right)^2\right] - \exp\left[-\left(X_{j-1}^{+}\right)^2\right],$$

$$L_{\rm erf}(E) = (E - E_{j-1})\left[\operatorname{erf}\left(X_{j-1}^{-}\right) - \operatorname{erf}\left(X_{j}^{-}\right)\right] + (E + E_{j-1})\left[\operatorname{erf}\left(X_{j}^{+}\right) - \operatorname{erf}\left(X_{j-1}^{+}\right)\right],$$

$$L_{\rm D}(E) = {\rm D}\left(X_{j}^{-}\right) - {\rm D}\left(X_{j-1}^{-}\right) + {\rm D}\left(X_{j-1}^{+}\right) - {\rm D}\left(X_{j}^{+}\right),$$

$$L_{\rm D_i}(E) = (E - E_{j-1})\left[{\rm D_i}\left(X_{j}^{-}\right) - {\rm D_i}\left(X_{j-1}^{-}\right)\right] + (E + E_{j-1})\left[{\rm D_i}\left(X_{j-1}^{+}\right) - {\rm D_i}\left(X_{j}^{+}\right)\right],$$

$$R_{\rm G}(E) = \exp\left[-\left(X_{j+1}^{-}\right)^2\right] - \exp\left[-\left(X_{j}^{-}\right)^2\right] + \exp\left[-\left(X_{j}^{+}\right)^2\right] - \exp\left[-\left(X_{j+1}^{+}\right)^2\right],$$

$$R_{\rm erf}(E) = (E - E_{j+1})\left[\operatorname{erf}\left(X_{j+1}^{-}\right) - \operatorname{erf}\left(X_{j}^{-}\right)\right] + (E + E_{j+1})\left[\operatorname{erf}\left(X_{j}^{+}\right) - \operatorname{erf}\left(X_{j+1}^{+}\right)\right],$$

$$R_{\rm D}(E) = {\rm D}\left(X_{j}^{-}\right) - {\rm D}\left(X_{j+1}^{-}\right) + {\rm D}\left(X_{j+1}^{+}\right) - {\rm D}\left(X_{j}^{+}\right),$$

$$R_{\rm D_i}(E) = (E - E_{j+1})\left[{\rm D_i}\left(X_{j}^{-}\right) - {\rm D_i}\left(X_{j+1}^{-}\right)\right] + (E + E_{j+1})\left[{\rm D_i}\left(X_{j+1}^{+}\right) - {\rm D_i}\left(X_{j}^{+}\right)\right].$$

The symbols $X_j^{\pm}$ and $X_{j\pm1}^{\pm}$ are defined by Eq. (B9).

## APPENDIX C: EXAMPLE DEMONSTRATING THE TDT MODEL

The TDT dispersion model is capable of describing band absorption more accurately than an approximation based on several discrete excitations. In this approach, the absorption band is represented by several TDT elements with characteristic excitation energies arranged in the ascending order.

Figure 16 presents an example of a band composed of four such elements; the corresponding parameters are listed in Table IX. The parameter values were chosen to illustrate various critical points of the dipole distribution function.

In this example, only a single parameter describing the broadening was assumed. Figure 16 shows the calculated response functions for three different values of this parameter. It follows that, for a given band broadening, it is necessary to choose an appropriate density of characteristic excitation energies, which typically correspond to the critical points associated with van Hove singularities.

Since the static value of the imaginary part of the response function diverges to $+\infty$, the present example does not describe a complete physical system. This divergence reflects the coupling of the absorption band to another part of the system, for example, to free carriers or to other bound excitations. In a non-conducting system, the remaining excitations would compensate for this divergence so that the total imaginary part of the susceptibility would vanish in the static limit.

For example, if the value of $M_1$ were smaller than approximately 0.13 for $B = 0.2$, the imaginary part would diverge to $-\infty$, i.e., the effective coupling to the rest of the system would have the opposite sign. To describe a non-conducting system fully within the framework of the present example, it would be necessary to constrain one of the coupling parameters by the others in order to ensure a zero static conductivity.

Finally, we note that the f-sum rule is identical for all response functions shown in Fig. 16 and is given by the sum of the transition-strength parameters,

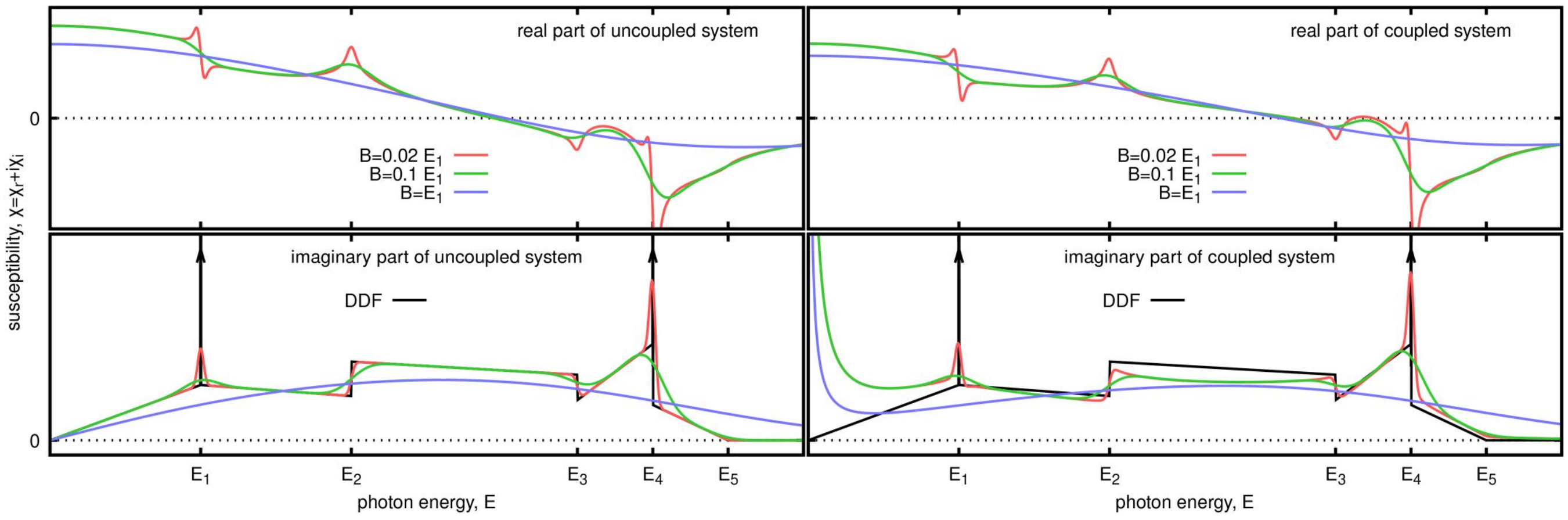

**FIG. 16.** Example illustrating the use of a triangle–delta–triangle (TDT) dispersion model consisting of four elements (see the parameters in Table IX). The susceptibility is calculated for three different values of the broadening parameter $B$. The left panels show the same response function as the right panels, but without coupling, i.e., with $M_j = 0$. The dipole distribution function (DDF) corresponds to the imaginary part of the susceptibility for $B = 0$ and $M_j = 0$.

**TABLE IX.** Dispersion parameters of the TDT model shown in Fig. 16. Parameters are given in arbitrary units.

| $j$ | $E_j$ | $N_{l,j}$ | $N_{d,j}$ | $N_{r,j}$ | $M_j$ |
|---|---|---|---|---|---|
| 1 | 2 | 1 | 0.1 | 2 | 1 |
| 2 | 4 | 2 | 0 | 8 | 2 |
| 3 | 7 | 8 | 0 | 2 | −2 |
| 4 | 8 | 5 | 1 | 2 | −1 |
| $E_0 = 0$, $E_5 = 9$, $B = 0.04, 0.2, 2$ | | | | | |

$$\int_0^\infty E\chi_i(E)\,dE = \sum_{k=l,d,r}\sum_{j=1}^{4} N_{k,j} = 31.1. \tag{C1}$$

Comparison of the imaginary part of the susceptibility for the uncoupled and coupled systems shows that the coupling parameters can significantly redistribute the transition strength and alter the static conductivity for the same dipole distribution.